\documentclass[prd,superscriptaddress,nofootinbib,amsfonts,notitlepage,longbibliography]{revtex4-1}
\usepackage[utf8]{inputenc}
\usepackage[T1]{fontenc}
\usepackage{hyperref}
\usepackage{graphicx,bm,amsmath,color,scalerel}
\usepackage{bbold}
\usepackage{wasysym}
\usepackage{verbatim}
\usepackage{amssymb}
\usepackage{epstopdf}

\usepackage{animate}
\usepackage{xmpmulti}
\usepackage{centernot}
\usepackage{multirow}
\usepackage{tikz}
\usepackage{comment}
\usepackage[normalem]{ulem} 
\makeatletter

\newcommand{\be}{\begin{equation}}
\newcommand{\ee}{\end{equation}}
\newcommand{\beq}{\begin{eqnarray}}
\newcommand{\eeq}{\end{eqnarray}}
\newcommand{\ba}{\begin{align}}
\newcommand{\ea}{\end{align}}

\begin{document}

\title{Time delays, choice of energy-momentum variables and relative locality in doubly special relativity}

\author{J.M. Carmona}
\email{jcarmona@unizar.es}
\affiliation{Departamento de F\'{\i}sica Te\'orica and Centro de Astropartículas y F\'{\i}sica de Altas Energ\'{\i}as (CAPA),
Universidad de Zaragoza, Zaragoza 50009, Spain}

\author{J.L. Cort\'es}
\email{cortes@unizar.es}
\affiliation{Departamento de F\'{\i}sica Te\'orica and Centro de Astropartículas y F\'{\i}sica de Altas Energ\'{\i}as (CAPA),
Universidad de Zaragoza, Zaragoza 50009, Spain}

\author{J.J. Relancio}
\email{jjrelancio@ubu.es}
\affiliation{Departamento de Física, Universidad de Burgos, 09001 Burgos, Spain}
\affiliation{Departamento de F\'{\i}sica Te\'orica and Centro de Astropartículas y F\'{\i}sica de Altas Energ\'{\i}as (CAPA),
Universidad de Zaragoza, Zaragoza 50009, Spain}

\author{M.A. Reyes}
\email{mkreyes@unizar.es}
\affiliation{Departamento de F\'{\i}sica Te\'orica and Centro de Astropartículas y F\'{\i}sica de Altas Energ\'{\i}as (CAPA),
Universidad de Zaragoza, Zaragoza 50009, Spain}
\begin{abstract}
Doubly Special Relativity (DSR) theories consider (quantum-gravity motivated) deformations of the symmetries of special relativity compatible with a relativity principle.  The existence of time delays for massless particles, one of their proposed phenomenological consequences, is a delicate question, since, contrary to what happens with Lorentz Invariance Violation (LIV) scenarios, they are not simply determined by the modification in the particle dispersion relation. While some studies of DSR assert the existence of photon time delays, in this paper we generalize a recently proposed model for time delay studies in DSR and show that the existence of photon time delays does not necessarily follow from a DSR scenario, determining in which cases this is so. Moreover, we clarify long-standing questions about the arbitrariness in the choice of the energy-momentum labels and the independence of the time delay on this choice, as well as on the consistency of its calculation with the relative locality paradigm of DSR theories. Finally, we show that the result for time delays is reproduced in models that consider propagation in a noncommutative spacetime.

\end{abstract}

\maketitle

\section{Introduction}
A quantum gravity theory (QGT) has been pursued during the last decades. This theory would reconcile Quantum Field Theory (QFT) and General Relativity (GR), and should be able to describe  particles with extremely-high energies, indispensable for studying the first instants of the universe.

Several attempts have been presented in order to avoid the inconsistencies between QFT and GR, such as string theory~\cite{Mukhi:2011zz,Aharony:1999ks,Dienes:1996du}, loop quantum gravity~\cite{Sahlmann:2010zf,Dupuis:2012yw}, 
causal dynamical triangulations \cite{Loll_2019}, or causal set \mbox{theory~\cite{Wallden:2013kka,Wallden:2010sh,Henson:2006kf}}. In most of these theories, a minimum length appears~\cite{Gross:1987ar,Amati:1988tn,Garay1995}, which is normally associated with the Planck length $\ell_P \sim 1.6\times 10^{-33}$\,cm, and, therefore, there is associated a high-energy scale.  Unfortunately, the aforementioned theories are not yet fully satisfactory in the sense that they still do not have well defined testable predictions, which might serve us as a guidance in building a definitive theory of quantum gravity.

Another way of thinking arose not long ago: instead of a fundamental QGT, one can consider a bottom-up approach in which residual effects of such a theory can be described. This opens up the possibility of testing phenomenological effects without an ultimate theory of quantum gravity (see~\cite{Addazi:2021xuf} for a recent review). In particular, one way of effectively describing these possible effects is to consider a modification of the kinematics of Special Relativity (SR). A crucial ingredient in this approach is the fate of the relativistic principle that characterizes SR, or in other words, of the Lorentz invariance of the theory. In particular, its breaking for high energies leads to Lorentz Invariance Violation (LIV) scenarios~\cite{Colladay:1998fq,Kostelecky:2008ts}, in which  there is a privileged observer. The main modification of the kinematics of SR is a modified dispersion relation, where usually new terms, proportional to the energy divided by the high-energy scale $\Lambda$ describing the SR deviation, are added to the expression of SR. 

On the other hand,  in Doubly/Deformed Special Relativity (DSR) theories~\cite{AmelinoCamelia:2008qg} the Lorentz symmetry is not violated, but deformed. In these scenarios, in addition to a deformation of the dispersion relation, there is a deformed (nonadditive) composition law for the momenta and, in order to save a relativity principle, some deformed Lorentz transformations for the one- and multi-particle systems. These relativistic deformed kinematics are usually constructed within the approach of Hopf algebras~\cite{Majid1994}, being $\kappa$-Poincaré kinematics~\cite{Lukierski:1991pn,Lukierski:1993df,Lukierski:1992dt,Lukierski:2002df} the most studied example in this context (see however~\cite{Carmona:2019fwf} for a geometrical interpretation of these kinematics). There are different bases of $\kappa$-Poincaré, i.e., different representations of the algebraic structure (see~\cite{KowalskiGlikman:2002jr} for some of them). It is important to note that the deformed Casimir depends on the chosen basis. For example,  in the so-called  ``classical basis'' of $\kappa$-Poincaré~\cite{Borowiec2010}, it is the same as in SR. However, the composition of momenta is always deformed in these kinematics, independently of the basis.      

Owing to the differences between LIV and DSR scenarios regarding Lorentz invariance, their phenomenological implications are rather different. While forbidden processes in SR are also not allowed in DSR (due to the relativistic principle present in this kind of theories), they can occur in a LIV scenario above some threshold energy~\cite{Mattingly:2005re,Liberati:2013xla}. Also, the threshold for allowed reactions are different. In LIV, the modification is of the order of $E^3/(m^2 \Lambda)$, where $E$ is one of the energies involved in the process in
our (Earth-based) laboratory frame, and $m$ is a mass that controls the corresponding threshold in SR. However, in DSR, due to cancellations of effects of the deformed dispersion relation and momentum composition law,  these modifications are of the order of $E/\Lambda$,  so that they are only relevant when the energy of the particles are of the order of the high-energy scale~\cite{Albalate:2018kcf,Carmona:2020whi,Relancio:2020mpa,Carmona:2021lxr}.  

Moreover, in both scenarios,  the possible existence of a time delay of photons has been proposed as an observable effect of a modification of special relativistic kinematics: due to an energy dependent velocity for massless particles, photons with different energies emitted simultaneously by a distant local source could be detected at different times. There is a clear consensus within the LIV community concluding that there is a time delay in this scenario~\cite{Ellis:2002in,RodriguezMartinez:2006ee,Jacob:2008bw}. Accordingly, the analysis of signals from different astrophysical sources (active galactic nuclei~\cite{HESS:2011aa,HESS:2019rhe}, gamma ray burst~\cite{MAGIC:2020egb,Du:2020uev}, and pulsars~\cite{Martinez:2008ki,Vasileiou:2013vra,MAGIC:2017vah}) have been used to put limits on, and in the case of some analyses~\cite{Xu:2016zxi,Li:2021tcw} identify hints of, an energy dependence of the velocity of propagation of photons due to a violation of Lorentz invariance (see~\cite{Addazi:2021xuf} for a more detailed discussion).

However, it is an open question whether one would expect time delays in the DSR framework. While there are several papers~\cite{Amelino-Camelia:2011ebd,Loret:2014uia,Rosati:2015pga,Mignemi:2016ilu} affirming the existence of such an effect,  some works point to the opposite~\cite{Carmona:2017oit,Carmona:2018xwm,Carmona:2019oph,Relancio:2020mpa}. The  reason behind the different contradictory answers to this question is the fact that the deformed implementation of the Poincaré transformations, in particular the translations between different frames, adds a new ingredient together with the velocity of propagation of photons in the discussion on time delays. This was considered  in~\cite{Amelino-Camelia:2011ebd}, where, besides a deformed dispersion relation, a noncommutativity of spacetime was present. A noncommutative spacetime was also the main ingredient in the analysis made in Ref.~\cite{Carmona:2017oit}.

In a recent work~\cite{frattulillo}, a systematic way for the study of time delays in DSR, based on postulating a representation of a deformed Poincaré algebra in canonical phase-space variables, was considered. Our aim in the present paper is to generalize this analysis to show that there is no contradiction between the different previous works claiming the presence or absence of time delays. In fact, both possibilities correspond to different cases of DSR, i.e., different bases of $\kappa$-Poincaré describe different physics. Moreover, we interpret different choices of energy-momentum labels as the result of canonical transformations. The identification of energy and momentum with the generators of translations has been a source of confusion in the literature~\cite{KowalskiGlikman:2002we,KowalskiGlikman:2002jr}. We will explicitly check that the expression of time delays is indeed independent of energy-momentum labels if they are identified with the momentum coordinates of the canonical phase space. This will allow us to single out specific choices for these labels in DSR theories. We will see that the choice of phase-space variables is also behind a non-conventional implementation of translations in the system of one particle. In contrast to what is commonly assumed, such nonstandard translations in the one-particle system are not a consequence of the relative locality paradigm of DSR, and in fact can be removed by an appropriate choice of the phase-space coordinates. We will show, however, that the analysis of time delays presented in the paper is consistent with the DSR formulation of the relative locality of interactions.

The paper is organized as follows. In Sec.~\ref{sec:rds}, we introduce the ingredients we will use in the computation of time delays: the most generic first-order deformation of the representation of translations and boosts in a canonical phase space, and the corresponding deformed dispersion relation. The parameter of the deformation will be identified with the inverse of a high-energy scale which characterizes the effects of new physics. We analyze in detail the computation of time delays in Sec.~\ref{sec:td}, including a discussion of the choice of the energy-momentum variables. In Sec.~\ref{sec:noncommutative}, we see that the same expression for time delays is obtained if instead of considering a commutative spacetime we use a noncommutative one, as it is done in some approaches of DSR. In Sec.~\ref{sec:MCLandRL} we study the consistency of the previous discussion of time delays with the main ingredients of DSR theories, a modified composition law and the relative locality of interactions. Finally, we end with the conclusions in Sec.~\ref{sec:conclusions}.

\section{Relativistic deformed symmetry}
\label{sec:rds}

From an algebraic point of view, DSR introduces a deformed Poincaré algebra characterized by a parameter in terms of which one can consider a power expansion of the deformation. Its inverse, which we will call $\Lambda$, has dimensions of energy, and we will associate it to a high-energy scale of new physics.

Let us consider the most general first-order deformation of the Poincaré transformations acting on the canonical phase space of one particle. These will be the ingredients that will enter the calculation of time delays that we will carry out in the following section. We denote by $E$, $P$, $N$ the generators of translations and boosts in a $1+1$ dimensional spacetime\footnote{The extension of the results to the case of $3+1$ dimensions is straightforward.} with coordinates ($x$,$t$). The canonical phase space is composed, together with the space-time coordinates, by its canonical conjugated momentum coordinates ($\Pi$, $\Omega$), such that $\{x,\Pi\}=1\;,\{t,\Omega\}=-1\;,\{x,\Omega\}=0$, and $\{t,\Pi\}=0$, where the $\{,\}$ operation stands for the Poisson bracket. The most general expression for the generators of the Poincaré transformations in terms of the canonical phase-space variables is\footnote{We have assumed the invariance of these expressions under the discrete transformation $\Pi\to -\Pi$, $P\to -P$, $x\to -x$, and $N\to -N$ in order to have an analogue in $1+1$ dimensions of the restrictions due to rotational invariance in the case of $3+1$ dimensions.} (here and in the following, equalities have to be understood neglecting $\mathcal{O}(1/\Lambda^2)$ terms):
\be
E \,=\, \Omega + \frac{a_1}{\Lambda} \,\Omega^2 + \frac{a_2}{\Lambda} \,\Pi^2\,, \quad\quad
P \,=\, \Pi + \frac{a_3}{\Lambda} \,\Omega \,\Pi\,, \quad\quad 
N \,=\, x \,\Omega - t \,\Pi + \frac{a_4}{\Lambda} \,x \,\Omega^2 + \frac{a_5}{\Lambda} \,x \,\Pi^2 - \frac{a_6}{\Lambda} \,t \,\Omega \,\Pi\,,
\label{eq:epn}
\ee
where the $a_i$ are adimensional coefficients determining the deformation of the Poincaré transformations in phase space. Eq.~\eqref{eq:epn} is a generalization of the model presented in \cite{frattulillo}.

The Casimir of this algebra will have the general form
\be
C \,=\, \Omega^2 - \Pi^2 + \frac{\alpha_1}{\Lambda} \,\Omega^3 + \frac{\alpha_2}{\Lambda} \,\Omega\,\Pi^2\,,
\label{Ca(Pi,Omega)}
\ee
where the coefficients $\alpha_1$, $\alpha_2$ are determined, in terms of the coefficients ($a_4$, $a_5$, $a_6$) in $N$, by the condition $\{N, C\}=0$. One obtains
\begin{align}
\{N, C\} \,=
\,\frac{(- 2 a_4 + 2 a_6 + 3 \alpha_1 + 2 \alpha_2)}{\Lambda} \,\Omega^2 \,\Pi + \frac{(- 2 a_5 + \alpha_2)}{\Lambda} \,\Pi^3 \,,
\end{align}
so that 
\be
\alpha_1 \,=\, \frac{(2 a_4 - 4 a_5 - 2 a_6)}{3} \quad\text{and}\quad \alpha_2 \,=\, 2 a_5\,.
\ee
The Casimir also determines the deformed energy-momentum relation, also known as dispersion relation. In the case of a massless particle ($C=0$), one finds 
\be
\Omega \,=\, \Pi - \frac{(\alpha_1+\alpha_2)}{2 \,\Lambda} \,\Pi^2  \,=\, \Pi - \frac{(a_4 + a_5 - a_6)}{3 \,\Lambda} \,\Pi^2 \,.
\label{energy-momentum}
\ee
One can also identify the deformed Poincaré algebra that these deformed generators satisfy,
\begin{align}
\{N, E\} \,=
\, \Pi + \frac{(a_6 + 2 a_1 + 2 a_2)}{\Lambda} \,\Omega \,\Pi  \,, \nonumber \\
\{N, P\} \,=
\, \Omega + \frac{(a_4 + a_3)}{\Lambda} \,\Omega^2 + \frac{(a_5 + a_3)}{\Lambda} \,\Pi^2  \,,
\end{align}
which is more naturally written in terms of the generators themselves,
\be
\{N, E\} \,= P + \frac{w_3}{\Lambda}EP  \,, \qquad
\{N, P\} \,= E + \frac{w_1}{\Lambda}E^2+\frac{w_2}{\Lambda}P^2  \,,
\label{eq:algebra-EPN}
\ee
where the coefficients $w_i$ are given by
\be
    w_1 = a_4 - a_1 + a_3 \,, \qquad w_2 = a_5 - a_2 + a_3 \,, \qquad w_3 = a_6 + 2 a_1 + 2 a_2 - a_3\,.
    \label{eq:algebra-EPN-coefficients}
\ee
Similarly, the Casimir can also be written in terms of $E$ and $P$,
\be
C \,=\, E^2 - P^2 + \frac{2 (w_1-2w_2-w_3)}{ 3\Lambda} \,E^3 + \frac{2 w_2}{\Lambda} \,E \,P^2\,.
\label{C(E,P)}
\ee

Note that the energy and momentum are denoted by $(\Omega,\Pi)$, so that one has to see the relation between $\Omega$ and $\Pi$, derived from the equation $C(\Omega,\Pi)=m^2$, as the particle dispersion relation which intervenes in the deformed kinematics, and it should not be confused with the relation between $E$ and $P$ that one could derive equating $C(E,P)$ to $m^2$. As we will remark in the following section, this is a source of confusion regarding the arbitrariness in the choice of energy-momentum labels for the physics under study.

\section{Calculation of time delays}
\label{sec:td}
Once the implementation of a deformed Poincaré algebra in the phase space of one particle has been set, it is possible to study its physical consequences for photon time delays, as we will do in the present section.

The deformed Casimir $C$ determines the propagation of particles in spacetime by identifying it with the generator of translations in a parameter $\tau$ along the worldline of the particle,
\be
\frac{dx}{d\tau} \,=\, \lambda (\tau) \,\{C, x\} \,, \quad\quad\quad
\frac{dt}{d\tau} \,=\, \lambda (\tau) \,\{C, t\} \,,
\ee
where $\lambda (\tau)$ is an arbitrary function implementing the invariance under reparametrizations of the worldline. The trajectory $x(t)$ will be a solution of 
\be
\frac{dx}{dt} \,=\, \frac{\{C, x\}}{\{C, t\}} \,=\, - \frac{\partial C/\partial \Pi}{\partial C/\partial\Omega}\,,\quad\quad
\frac{d\Omega}{d\tau} \,=\, \frac{d\Pi}{d\tau} \,=\, 0\,, \quad\quad
C(\Omega, \Pi)\,=\, m^2\,.
\ee
Using the dispersion relation for a massless particle, Eq.~\eqref{energy-momentum}, which is the relevant one for the discussion of photon time delays, one obtains that the velocity of propagation for photons is
\be
v=\frac{dx}{dt} \,=\, 1 - \frac{(\alpha_1+\alpha_2)}{\Lambda} \,\Pi  \,=\, 1 - \frac{2\,(a_4+a_5-a_6)}{3\, \Lambda} \,\Pi \,,
\label{dx/dt}
\ee
a result that could have been derived directly from the energy-momentum relation identifying the velocity with $d\Omega/d\Pi$.

In order to discuss time delays, we consider the propagation of two photons, one of high energy, whose trajectory will be characterized by Eq.~\eqref{dx/dt}, and one with low energy, that propagates with $dx/dt\approx 1$, as in SR. We will define the time delay as the difference in the detection times of the two photons if their emission was simultaneous.

\begin{figure}[thbp]
    \centering
    \begin{minipage}{0.32\textwidth}
        \centering
        \includegraphics[width=0.8\textwidth]{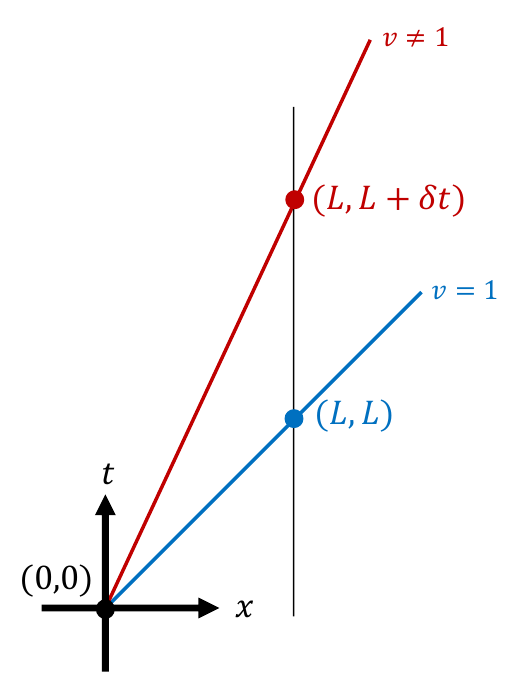}
        \caption{In LIV, only one observer is necessary for the calculation of time delays. An observer at the emission is able to assign coordinates to the detection of both photons.}
        \label{fig:LIV}
    \end{minipage}%
    \hfill
    \begin{minipage}{0.66\textwidth}
        \centering
        \includegraphics[width=0.39\textwidth]{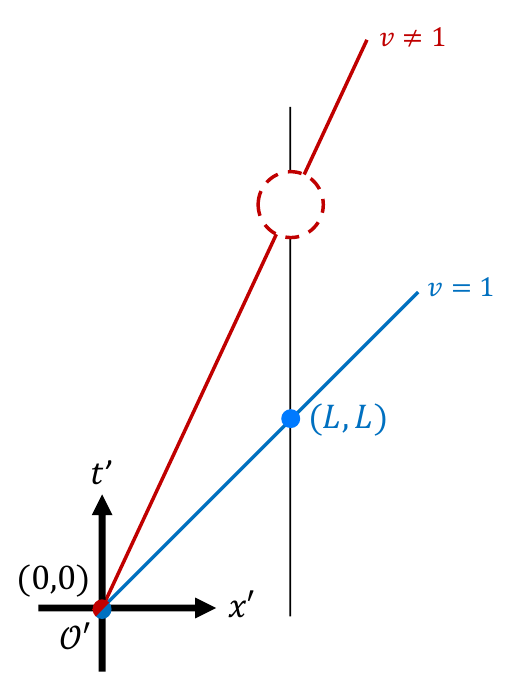}
        \includegraphics[width=0.39\textwidth]{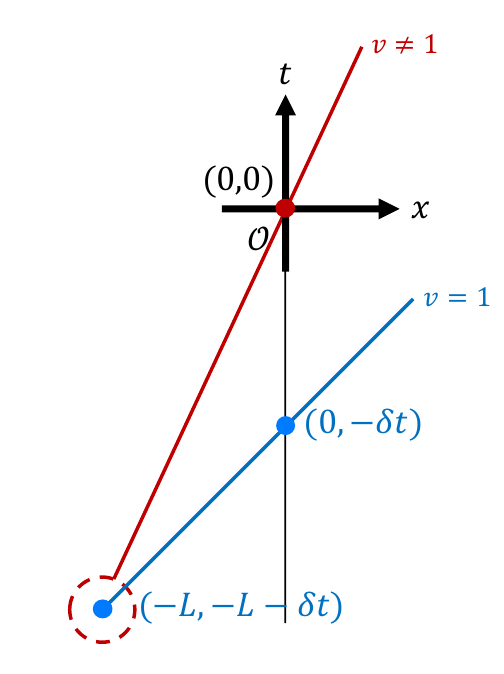}
        \caption{In DSR, due to relative locality, the observer $\mathcal O'$ (left figure) sees the emission of the high-energy photon as local, but not the detection. The opposite behavior is shown for $\mathcal O$ (right figure), which sees the detection of the high-photon as local, but not the emission. The emission and detection of the low-energy photon is local for every observer.}
        \label{fig:DSR}
    \end{minipage}
\end{figure}

In the case of a LIV scenario, one only needs to consider a single observer to discuss time delays. If the two photons are emitted at the origin of the space-time coordinates, the low-energy photon is detected at coordinates $(L,L)$, and the high-energy photon, propagating at speed $v$, at coordinates $(L,L+\delta t)$, where $\delta t$ is the time delay, which is equal to  $L[(1/v)-1]$ (Fig.~\ref{fig:LIV}).

In contrast, the analysis of time delays is much more subtle in a DSR scenario. The reason is the principle of relative locality~\cite{AmelinoCamelia:2011bm}, which states that in DSR, for a given interaction (e.g., either the emission or the detection of the high-energy photon), there exists only one observer\footnote{We are here identifying observers related by a Lorentz transformation, so that different observers are related by a translation.} which sees this interaction as local. For this `local' observer, the interaction takes place at the origin of the space-time coordinates. The nonlocality of all other interactions is nevertheless negligible when all the intervening momenta are sufficiently small compared with the scale $\Lambda$. We will analyze in more detail the connection between the present calculation of time delays and relative locality in Sec.~\ref{sec:MCLandRL}.

Relative locality makes the presence of two observers, one which is local to the emission of the high-energy photon, and another one which is local to its detection, necessary in order to compute the time delay. Let us call $\mathcal O$ the observer for which the detection of the high-energy photon happens at its origin of space-time coordinates, $(x,t)=(0,0)$ (Fig.~\ref{fig:DSR}, right), and $\mathcal O'$ the observer for which the emission of the high-energy photon takes place at $(x',t')=(0,0)$ (Fig.~\ref{fig:DSR}, left). 

If the source and the detector are at relative rest, the coordinates of the points of the photon worldline assigned by each observer will be related by a translation with parameters $(\epsilon_1, \epsilon_0)$ that we need to determine. Since the Poisson brackets of $E$ or $P$ with $\{P,x\}$, $\{P,t\}$, $\{E,x\}$ and $\{E,t\}$ are zero, one can compute the finite transformation as
\be
    x'\,=\,x- \epsilon_1 \{P, x\} + \epsilon_0 \{E, x\} \,, \quad\quad
    t'\,=\,t - \epsilon_1\{P, t\} + \epsilon_0 \{E, t\} \,.
    \label{eq:coordtransf}
\ee
As for both observers the wordline of the high-energy photon contains their corresponding origin of spacetime, its trajectory will be $x=vt$ for observer $\mathcal{O}$, and $x'=vt'$ for observer $\mathcal{O'}$, where $v$ is given by Eq.~\eqref{dx/dt}. This imposes a relation between the parameters of the translation,
\be
    \epsilon_0 \left[v\{E,t\} - \{E,x\}\right] \,=\, \epsilon_1 \left[v\{P,t\} - \{P,x\}\right]\,.
\ee
Using that
\be
    \{E,t\}=1+\frac{2a_1}{\Lambda}\Omega\,,\quad \{E,x\}=-2\frac{a_2}{\Lambda}\Pi\,,\quad \{P,t\}=\frac{a_3}{\Lambda}\Pi\,,\quad \text{and}\quad \{P,x\}=-1-\frac{a_3}{\Lambda}\Omega\,,
\ee
and the energy-momentum relation, Eq.~\eqref{energy-momentum}, one gets
\be
    \frac{\epsilon_0}{\epsilon_1} = 1+ \frac{2(a_4+a_5-a_6)}{3\Lambda}\,\Pi - \frac{2(a_1+a_2-a_3)}{\Lambda} \,\Pi \,.
    \label{eq:parameters}
\ee

In order to determine these parameters, let us consider the simultaneous emission of a low-energy photon, that is, emitted at $(x',t')=(0,0)$, whose trajectory is characterized by $dx/dt\approx 1$. The nonlocality of the emission and detection of this photon can be neglected for all observers if one assumes that these are low-energy processes. Observer $\mathcal{O}$ defines the detection of the high-energy photon at $x=0$. Disregarding the size of the detector and since any relative locality effects between $\mathcal{O}$ and an observer who detected the low-energy photon at the origin of its space-time coordinates is negligible, the detection of the low-energy photon will also happen at $x=0$ for $\mathcal{O}$. The coordinates of the detection of the low-energy photon will be then $(x,t)=(0,-\delta t)$, where $\delta t$ is the time delay of the high-energy photon with respect to the one of low energy. This detection will correspond to a point $(x',t')=(L,L)$ for $\mathcal{O'}$ (see Fig.~\ref{fig:DSR}). The relation between the coordinates assigned by both observers to the worldline of the low-energy photon is then
\be
    x'\,=\,x + L \,, \quad\quad
    t'\,=\,t + (L+\delta t)\,.
    \label{eq:coordtransf-low}
\ee
This gives the values of the parameters of the translation in Eq.~\eqref{eq:coordtransf}, $\epsilon_1=L$, $\epsilon_0=L+\delta t$, and then,
\be
   \frac{\epsilon_0}{\epsilon_1} = 1+ \frac{\delta t}{L} \,.
    \label{eq:parameters-low}
\ee
From Eq.~\eqref{eq:parameters} we get
\be 
    \delta t \,=\, L \left[ \frac{2(a_4+a_5-a_6)}{3\Lambda}\,\Pi - \frac{2(a_1+a_2-a_3)}{\Lambda} \,\Pi\right]\,.
    \label{delta t}
\ee
We see that the time delay is a sum of two contributions. The first one, due to the momentum dependence of the velocity of propagation of photons, can be read from the equation for the trajectory, Eq.~\eqref{dx/dt}, and the second one, due to the nontrivial translations relating the two observers $\mathcal{O}$ and $\mathcal{O'}$.

One will have a cancellation of the two contributions in \eqref{delta t}, and then no observable consequences at the level of time delays, when the deformations of the boost and space-time translation generators are such that
\be
a_4+a_5-a_6 \,=\, 3 (a_1+a_2-a_3)\,.
\label{no-time-delay}
\ee

\subsection{Independence of the choice of energy-momentum variables}
\label{sec:choice}
In the previous study, we have identified the energy-momentum variables with the momentum coordinates $(\Pi,\Omega)$ of the canonical phase space; however, there is not a unique choice for these coordinates, since one can consider a canonical change of phase-space variables $(x,t,\Pi,\Omega) \to (\overline{x},\overline{t},\overline{\Pi},\overline{\Omega})$, with 
\be
\overline{\Omega} \,=\, \Omega + \frac{\delta_1}{\Lambda} \,\Omega^2 + \frac{\delta_2}{\Lambda} \,\Pi^2\,,
\quad\quad
\overline{\Pi} \,=\, \Pi + \frac{\delta_3}{\Lambda} \,\Omega \,\Pi\,,
\quad\quad
\overline{t} \,=\, t - \frac{2 \delta_1}{\Lambda} \,t \Omega + \frac{\delta_3}{\Lambda} \,x \,\Pi\,,
\quad\quad 
\overline{x} \,=\, x + \frac{2 \delta_2}{\Lambda} \,t \,\Pi - \frac{\delta_3}{\Lambda} \,x \,\Omega\,.
\label{canonicaltr}
\ee
The generators of the deformed Poincaré algebra acting on the new phase-space coordinates are then
\be
E \,=\, \overline{\Omega} + \frac{\overline{a}_1}{\Lambda} \,\overline{\Omega}^2 + \frac{\overline{a}_2}{\Lambda} \,\overline{\Pi}^2\,, \quad\quad
P \,=\, \overline{\Pi} + \frac{\overline{a}_3}{\Lambda} \,\overline{\Omega} \,\overline{\Pi}\,, \quad\quad 
N \,=\, \overline{x} \,\overline{\Omega} - \overline{t} \,\overline{\Pi} + \frac{\overline{a}_4}{\Lambda} \,\overline{x} \,\overline{\Omega}^2 + \frac{\overline{a}_5}{\Lambda} \,\overline{x} \,\overline{\Pi}^2 - \frac{\overline{a}_6}{\Lambda} \,\overline{t} \,\overline{\Omega} \,\overline{\Pi}\,,
\ee
with
\begin{gather}
\overline{a}_1 \,=\, a_1 - \delta_1\,, \quad
\overline{a}_2 \,=\, a_2 - \delta_2\,, \quad
\overline{a}_3 \,=\, a_3 - \delta_3\,, \\
\overline{a}_4 \,=\, a_4 - \delta_1 + \delta_3\,, \quad
\overline{a}_5 \,=\, a_5 - \delta_2 + \delta_3\,, \quad
\overline{a}_6 \,=\, a_6 + 2 \delta_1 + 2 \delta_2 - \delta_3 \nonumber\,.
\end{gather}
Hence one can check that
\be
\delta \overline t \,=\, L\;\frac{2(\overline a_4+\overline a_5-\overline a_6-3\overline a_1-3\overline a_2+3\overline a_3)}{3\Lambda}\,\overline \Pi \,=\, L\;\frac{2(a_4+a_5-a_6-3a_1-3 a_2+3 a_3)}{3\Lambda}\,\Pi \,=\, \delta t \,.
\ee
From this result one concludes that the value of the time delay, i.e. Eq~\eqref{delta t}, is independent of the choice of energy-momentum variables.

\subsection{Coordinate-independent expression of the time delay}
We can use this freedom of choice of the energy-momentum variables to simplify the computation of the time delay. For example, from Eq.~\eqref{dx/dt}, one can check that the condition to have a momentum-independent velocity for massless particles is $(\overline a_4 + \overline a_5 - \overline a_6) = 0$. Therefore, when performing a change of phase-space variables such that $\delta_1 + \delta_2 - \delta_3 = (1/3) (a_4 + a_5 - a_6)$, the time delay will have a contribution only from the nontrivial translations. On the other hand, from Eq.~\eqref{delta t}, we see that the condition to have trivial translations is $(\overline a_1+ \overline a_2- \overline a_3)=0$. Then, when performing a change of phase-space variables such that $\delta_1 + \delta_2 - \delta_3 = (a_1 + a_2 - a_3)$, the time delay  will be due exclusively to the momentum dependence of the velocity.

Let us use a choice of energy-momentum variables such that they coincide with the generators of space-time translations,  $(\overline\Omega,\overline\Pi)=(E, P)$, and so
\be
\overline{t} \,=\, t - \frac{2 a_1}{\Lambda} \,t \,\Omega + \frac{a_3}{\Lambda} \,x \,\Pi\,, \quad\quad\quad
\overline{x} \,=\, x + \frac{2 a_2}{\Lambda} \,t \,\Pi - \frac{a_3}{\Lambda} \,x \,\Omega\,.
\ee
With this choice of coordinates, the space-time translations are not deformed, so the time delay is only due to the momentum dependence of the velocity, i.e. $\delta \overline t=L[1/\overline v-1]$. Moreover, the velocity can be read directly from the Casimir in terms of $E$ and $P$, 
\be
\overline v = \frac{d\overline{x}}{d\overline{t}} \,=\, \frac{d\overline\Omega}{d\overline\Pi} \,=\, \frac{dE}{dP}\bigg |_{C=0}.
\ee
Then the time delay, Eq.~\eqref{delta t}, can also be written in terms of the relation between $E$ and $P$, which does not depend on the choice of phase-space variables.
\be
\delta t \,=\, \delta \overline t \,=\, L\left[\left(\frac{dE}{dP}\bigg |_{C=0}\right)^{-1} -1 \right] \,=\, L \, \frac{2(w_1+w_2-w_3)}{3 \Lambda} \,P\,.
\ee
Therefore, the condition of absence of time delays can be written as a condition over the algebra of $(E,P,N)$,
\be
w_1 + w_2 - w_3 \,=\, 0\,,
\label{no-time-delay-2}
\ee
which is independent of the choice of phase-space variables. Without losing generality, then, one can say that there will be no time delay as long as the Casimir expressed in term of $E$ and $P$ is such that the solution of the equation $C(E,P)=0$ is the same as in SR, i.e., $E=P$.\footnote{Note that this is not the same as saying that the energy is equal to the modulus of the momentum for a photon, since the energy and momentum of the photon are $\Omega$ and $\Pi$, respectively, and, in general, they do not coincide with the generators of the space-time translations $E$ and $P$.}

We can now show some examples of different bases of $\kappa$-Poincaré, for which the previous condition is satisfied:
\begin{itemize}
\item Classical basis~\cite{Borowiec2010}

In this basis, the Casimir is unmodified with respect to the SR case. Therefore, $E=P$ trivially for photons. 

\item Magueijo-Smolin basis~\cite{Magueijo:2001cr}

The Casimir in this basis is 
\be
C\,=\,\frac{E^2-P^2}{\left(1-E/\Lambda\right)^2}\,,
\ee
for which $E=P$ for massless particles.

\item DCL1 basis~\cite{Carmona:2019vsh}  

The Casimir in this case is given by
\be
C\,=\,\frac{E^2-P^2}{\left(1-E/\Lambda\right)}\,.
\ee
Again, $E=P$ for massless particles.

\end{itemize}

\subsection{Change of basis of \texorpdfstring{$\kappa$}{k}-Poincaré}
Let us notice that a very different situation arises if one, instead of making a change of phase-space coordinates, now changes the deformed generators of space-time translations, i.e., if one changes the basis of $\kappa$-Poincaré. This implies that we change $(E,P,N) \to (\overline{E},\overline{P},N)$, with
\be
\overline{E} \,=\, E + \frac{\Delta_1}{\Lambda} \,E^2 + \frac{\Delta_2}{\Lambda} \,P^2\,, 
\quad\quad\quad
\overline{P} \,=\, P + \frac{\Delta_3}{\Lambda} \,E \,P\,.
\ee
Then we have a new deformed algebra given by
\be
\{N, \overline{P}\} \,=\, \overline{E} + \frac{\overline{w}_1}{\Lambda} \,\overline{E}^2 + \frac{\overline{w}_2}{\Lambda} \,\overline{P}^2\,, 
\quad\quad\quad
\{N, \overline{E}\} \,=\, \overline{P} + \frac{\overline{w}_3}{\Lambda} \,\overline{E} \,\overline{P}\,,
\ee
with
\be
\overline{w}_1 \,=\, w_1 - \Delta_1 + \Delta_3 \,, \quad\quad\quad
\overline{w}_2 \,=\, w_2 - \Delta_2 + \Delta_3 \,, \quad\quad\quad
\overline{w}_3 \,=\, w_3 + 2 \Delta_1 + 2 \Delta_2 - \Delta_3 \,,
\ee
and, as a consequence, the result for time delays is also modified,
\be
\delta \overline{t} \,=\, L \, \frac{2(\overline{w}_1+\overline{w}_2-\overline{w}_3)}{3 \Lambda} \,\Pi \,=\, L \, \frac{2(w_1+w_2-w_3)}{3 \Lambda} \,\Pi - L \,\frac{2(\Delta_1+\Delta_2-\Delta_3)}{\Lambda} \,\Pi \,=\, \delta t - L \,\frac{2(\Delta_1+\Delta_2-\Delta_3)}{\Lambda} \,\Pi\,.
\ee
Different bases of $\kappa$-Poincaré, therefore, correspond to different physical models going beyond SR.

\section{Time delays in a noncommutative spacetime}
\label{sec:noncommutative}

Some works have considered a noncommutative spacetime as the physical spacetime where trajectories of particles should be defined~\cite{Carmona:2017oit,Carmona:2019oph,Amelino-Camelia:2011ebd,Mignemi:2016ilu}. In particular, the $\kappa$-Minkowski spacetime is associated naturally to a $\kappa$-Poincaré deformation of the algebra of special relativity~\cite{KowalskiGlikman:2002we}, which has been extensively studied in connection with DSR theories. In this section we will examine how the previous discussion of time delays is modified if one considers that the propagation and detection of photons are described in such a spacetime.

In order to introduce this noncommutative spacetime, let us consider the deformation of Poincaré transformations acting on momentum space which include, together with the generator of boosts $N$, the deformed generators of translations ($X$, $T$) acting on the momentum variables ($\Pi$, $\Omega$).
Such a deformation is necessarily of the form\footnote{Note that the first parenthesis of~\eqref{lie} is  a definition of the scale $\Lambda$ used in the previous sections.}
\be
\{T, X\} \,=\, \frac{1}{\Lambda} \,X\,, \quad\quad\quad
\{N,X\} \,=\, T\,, \quad\quad\quad
\{N, T\} \,=\, X + \frac{1}{\Lambda} \,N\,.
\label{lie}
\ee

A general expression for these deformed generators which is linear in the space-time coordinates is
\be
X \,=\, x + \frac{b_1}{\Lambda} \,x \,\Omega + \frac{b_2}{\Lambda} \,t \,\Pi\,, \quad\quad\quad 
T \,=\, t + \frac{b_3}{\Lambda} \,x \,\Pi + \frac{b_4}{\Lambda} \,t \,\Omega\,.
\label{gen_translation_momentum}
\ee
The coefficients $b_i$ are determined by the Lie algebra; using
\begin{align}
\{N, X\} \,=& 
\, t + \,\frac{(b_1+b_2-2a_5)}{\Lambda} \,x \,\Pi + \frac{(b_1+b_2+a_6)}{\Lambda} \,t \,\Omega \,,
\nonumber \\
\{N, T\} \,=&
\, x + \frac{(b_3+b_4+2a_4)}{\Lambda} \,x \,\Omega + \frac{(b_3+b_4-a_6)}{\Lambda} \,t \,\Pi \,,
\nonumber \\
\{X, T\} \,=&
\, \frac{(b_1+b_3)}{\Lambda} \,x \,,
\end{align}
and Eqs.~\eqref{lie}-\eqref{gen_translation_momentum}, we get
\be
b_1\,=\,\frac{2a_4+2a_5+a_6}{3}-1\,, \quad
b_2\,=\,\frac{-4a_4+2a_5-2a_6}{3}+1\,,
\quad
b_3\,=\,\frac{-2a_4-2a_5-a_6}{3}\,,
\quad
b_4\,=\,\frac{-2a_4+4a_5+2a_6}{3}\,.
\label{b1b2b3b4}
\ee

One can make a model in which the physical space-time coordinates are ($X$,$T$) instead of the canonical ones ($x$,$t$) used in the previous section. From Eq.~\eqref{gen_translation_momentum}, we find
\be
    \frac{dX}{dt} \,=\,\frac{dx}{dt}+\frac{b_1}{\Lambda}\frac{dx}{dt}\Omega+\frac{b_2}{\Lambda}\Pi\,, \quad\quad\quad
    \frac{dT}{dt} \,=\,1+\frac{b_3}{\Lambda}\frac{dx}{dt}\Pi+\frac{b_4}{\Lambda}\Omega\,,
\ee
and then, merging both equations and expanding until order $1/\Lambda$, we get that the velocity in the noncommutative spacetime is given by
\begin{align}
    \frac{dX}{dT} &\,=\, \frac{dx}{dt}+ \frac{(b_1+b_2-b_3-b_4)}{\Lambda}\Pi  \notag \\ &\,=\, \left(1 - \frac{2\,(a_4+a_5-a_6)}{3\, \Lambda}\Pi \right)+  \frac{2\,(a_4+a_5-a_6)}{3\, \Lambda}\Pi  \,=\, 1  \,,
\end{align}
where in the next-to-last step we have substituted the values of $(b_1,b_2,b_3,b_4)$, given by Eq~\eqref{b1b2b3b4}. Then, we get that the velocity of a massless particle in this noncommutative spacetime is  simply $(dX/dT)=1$.

A point with coordinates $(x,t)=(0,0)$ has coordinates $(X,T)=(0,0)$. Then, the detection of the high-energy photon happens at $(X,T)=(0,0)$ for observer $\mathcal O$, and its emission at $(X',T')=(0,0)$ for observer $\mathcal O'$. The worldline for the photon is then $X=T$ for observer $\mathcal O$, and $X'=T'$ for observer $\mathcal O'$. Imposing once more that one can go from observer $\mathcal O$ to observer $\mathcal O'$ by a translation with parameters $(\epsilon_1,\epsilon_0)$, one has
\be
X' \,=\, X + \epsilon_0 \{E, X\} - \epsilon_1 \{P, X\} \,, \quad\quad
T' \,=\, T + \epsilon_0 \{E, T\} - \epsilon_1 \{P, T\} \,. 
\ee
Applying the translation to the worldline $X=T$ of the observer $\mathcal O$, one has to find the worldline $X'=T'$ for the observer $\mathcal O'$. This implies the relation
\be
\epsilon_0 \left[\{E,T\} - \{E,X\}\right] \,=\, \epsilon_1 \left[\{P,T\} - \{P,X\}\right]\,.
\ee

Combining the expressions of the noncommutative space-time coordinates, in terms of the canonical phase-space variables in \eqref{gen_translation_momentum}, with the expressions of ($E,P$), in terms of the momentum variables ($\Omega,\Pi$) in \eqref{eq:epn}, one has
\begin{align}
&\{E,T\} \,=\, 1 + \frac{(b_4 + 2 a_1)}{\Lambda} \Omega\,, \quad\quad
\{E,X\} \,=\, \frac{(b_2 - 2 a_2)}{\Lambda} \Pi\,, \nonumber \\
&\{P,T\} \,=\, \frac{(-b_3 + a_3)}{\Lambda} \Pi\,, \quad\quad
\{P,X\} \,=\, -1 - \frac{(b_1+a_3)}{\Lambda} \Omega\,.
\end{align}
Using once more the energy-momentum relation, Eq.~\eqref{energy-momentum}, one finds
\be
\frac{\epsilon_0}{\epsilon_1} \,=\, 1 + \frac{(b_1+b_2-b_3-b_4)}{\Lambda} \Pi - \frac{2(a_1+a_2-a_3)}{\Lambda} \Pi \,=\, 1 + \frac{2(a_4+a_5-a_6)}{3 \Lambda} \Pi - \frac{2(a_1+a_2-a_3)}{\Lambda} \Pi \,. 
\ee
The parameters of the translation between observers $\mathcal O$ and $\mathcal O'$ defined in a noncommutative spacetime satisfy exactly the same relation, Eq.~\eqref{eq:parameters}, obtained previously. This is just a consequence that a point with coordinates $(x,t)=(0,0)$ is a point with coordinates $(X,T)=(0,0)$, which is what defines the two observers, $\mathcal O$ and $\mathcal O'$, in the model of time delays. Since for the trajectory of the low-energy photon, which allows one to define the time delay, there is no distinction between commutative and noncommutative coordinates, the result for the time delay $\delta T$  coincides with the result for $\delta t$, Eq.~\eqref{delta t}, obtained in Sec.~\ref{sec:td}.

\section{Modified composition law and relative locality in DSR}
\label{sec:MCLandRL}

DSR contains an essential feature in comparison with SR and LIV: a nonadditive composition law for momenta, which can be seen as a reflection of a curved momentum space. As a consequence of this feature, a relative locality of interactions arises, since translations in spacetime must be deformed accordingly~\cite{AmelinoCamelia:2011bm}. In this section we will explore how the properties of a modified composition law and relative locality are related to the discussion of time delays of the preceding sections. 

\subsection{Modified composition law}
\label{sec:MCL}

In all the previous discussion we have considered a deformation of the Poincaré transformations acting on the one-particle system. This is an ingredient required to study the presence of time delays in the detection of photons of different energies emitted simultaneously by a local source as a possible consequence of the relativistic deformation of SR. But the physical content of a relativistic deformation is in the nontrivial step going from one-particle to multi-particle states. This is manifest in the algebraic framework at the level of the coalgebra involved in the Hopf algebra which characterizes the deformed relativistic symmetry~\cite{Lukierski:1992dt}. In the purely kinematic perspective, a DSR theory has a non-symmmetric composition law of momenta as the ingredient defining the deformation of the kinematics. We will now see how this key feature of DSR is related with the ingredients used in the calculation of the time delay.

Let us consider a translation in momentum space, with generators ($X$,$T$) and parameters \mbox{($\pi$, $\omega$)}. Using Eq.~\eqref{gen_translation_momentum}, we get
\begin{align}
\Pi' - \Pi\,=&\, \pi \{X, \Pi\} - \omega \{T, \Pi\} \,=
\, \pi + \frac{b_1}{\Lambda} \,\pi \,\Omega - \frac{b_3}{\Lambda} \,\omega \,\Pi\,, 
\nonumber \\
\Omega'-\Omega \,=& \,\pi \{X, \Omega\} - \omega \{T, \Omega\} \,=
\, \omega - \frac{b_2}{\Lambda} \,\pi \,\Pi + \frac{b_4}{\Lambda} \,\omega \,\Omega\,.
\end{align}
Defining a composition of two momentum variables $q=(q_1,q_0)=(\Pi, \Omega)$ and $p=(p_1,p_0)=(\pi, \omega)$ as the momentum variable that results from the previous translation 
\be
(p\oplus q)_1 \,=\, p_1 + q_1 + \frac{\gamma_1}{\Lambda} \,p_0 \,q_1 + \frac{\gamma_2}{\Lambda} \,p_1\,q_0 \,\doteq\, \Pi'\,,\quad\quad
(p\oplus q)_0 \,=\, p_0 + q_0 + \frac{\beta_1}{\Lambda} \,p_0\,q_0 + \frac{\beta_2}{\Lambda} \,p_1\,q_1 \,\doteq\, \Omega' \,,
\label{eq:composition}
\ee
one has 
\be
\gamma_1 \,=\, - b_3  \,,\quad\quad
\gamma_2 \,=\, b_1  \,,\quad\quad
\beta_1 \,=\, b_4  \,,\quad\quad
\beta_2 \,=\, - b_2  \,.\quad\quad
\label{eq:gb}
\ee
The coefficients defining the deformed composition of momenta $(\beta_1,\beta_2,\gamma_1,\gamma_2)$ are in one-to-one correspondence with the coefficients ($b_1$, $b_2$, $b_3$, $b_4$). This way, the deformed composition of momenta determines the coefficients ($a_4$, $a_5$, $a_6$) of the deformed boost generator using Eq.~\eqref{b1b2b3b4}. The latter also determines the modified energy-momentum relation, or dispersion relation, given by Eq.~\eqref{energy-momentum}. In fact one has 
\be
\alpha_1 \,=\, \frac{2a_4-4a_5-2a_6}{3} \,=\, - b_4 \,=\, - \beta_1\,,
\quad\quad\quad
\alpha_2 \,=\, 2 a_5 \,=\, b_1 + b_2 - b_3 \,=\, \gamma_1 + \gamma_2 - \beta_2\,,
\ee
which are the golden rules of a relativistic kinematics derived in a previous work~\cite{Carmona:2012un} from the compatibility of the deformed composition of momenta and the implementation of Lorentz transformations in a two-particle system.

A DSR theory always has a modified composition law of momenta, giving rise to non-standard conservation laws in particle processes, and therefore, to possible observable effects beyond time delays. This fact is what makes a DSR model without time delays different from standard SR (see however~\cite{Arzano:2019toz,Lobo:2020qoa,Lobo:2021yem} for recent works discussing a possible effect of DSR in the lifetime of particles).  

\subsection{Relative locality}
\label{sec:RL}

A modified composition law leads to a loss of absolute locality in canonical spacetime~\cite{AmelinoCamelia:2011bm,Carmona:2017cry,Carmona:2018xwm,Carmona:2019vsh,Carmona:2019oph,Gubitosi:2019ymi,Carmona:2021gbg}. Qualitatively, since the total momentum, which is conserved in an interaction, is a nonlinear composition of momenta, translations (generated by the total momentum) of the worldlines participating in the interaction are not constant displacements (as it would be the case if the total momentum were a linear sum of momenta) but momentum-dependent displacements: a local interaction for one observer is then seen as nonlocal for a translated observer. 

A realization of the relative locality of interactions in DSR is the action formulation of an interaction that was introduced in Ref.~\cite{AmelinoCamelia:2011bm}. The action is of the form
\begin{equation}
    S\,=\,\sum_i S_i^\text{free}+S^\text{int}\,,
    \label{eq:action}
\end{equation}
where $i$ labels the particles (both \emph{in}coming and \emph{out}going) participating in the interaction. The free part is
\begin{equation}
    S_i^\text{free}\,=\,\int d\tau \,(x_i \dot{\Pi}_i - t_i \dot{\Omega}_i + \mathcal{N}_i C_i)\,,
\end{equation}
where the integral extends from $-\infty$ to $0$ (if we set the interaction to take place at $\tau=0$ for each of the particles) in the case of the \emph{in} particles, and from $0$ to $+\infty$ in the case of the \emph{out} particles. In the previous expression, $\mathcal{N}_i$ are Lagrange multipliers that impose the modified dispersion relations $C_i=0$. The interacting part is
\begin{equation}
    S^\text{int}\,=\,z_1[\mathcal{P}^\text{\emph{in}}(0)-\mathcal{P}^\text{\emph{out}}(0)] - z_0[\mathcal{E}^\text{\emph{in}}(0)-\mathcal{E}^\text{\emph{out}}(0)]\,,
    \label{eq:RLint}
\end{equation}
where $\mathcal{E}$ and $\mathcal{P}$ are the total energy and momentum which are obtained from the modified composition law of the incoming or outgoing momenta, and $(z_1,z_0)$ are again Lagrange multipliers.

From the variational principle, one obtains that the \emph{in} and \emph{out} worldlines are characterized by constant momenta, independent of $\tau$, and gets the space-time coordinates of their corresponding end or starting points,  
\begin{equation}
    t_i(0)\,=\,-z_1\frac{\partial \mathcal{P}}{\partial \Omega_i} + z_0\frac{\partial \mathcal{E}}{\partial \Omega_i}\,=\,
-z_1 \{\mathcal{P},t_i\} + z_0 \{\mathcal{E},t_i\}\,, \quad \quad
    x_i(0)\,=\,z_1\frac{\partial \mathcal{P}}{\partial \Pi_i} - z_0\frac{\partial \mathcal{E}}{\partial \Pi_i}\,=\,
-z_1 \{\mathcal{P},x_i\} + z_0 \{\mathcal{E},x_i\}\,.  
    \label{eq:RLpoint}
\end{equation}
The set of momenta and the values of $(z_1,z_0)$ determine the end or starting points of the \emph{in} and \emph{out} worldlines, respectively, and the interaction is seen as local only for the observer which establishes the origin of space-time coordinates at the interaction vertex (corresponding to $z_1=z_0=0$), as shown in Eq.~\eqref{eq:RLpoint}. An observer who assigns a different value of the pair $(z_1,z_0)$ to the interaction is related with the local observer by a translation generated by the total energy and momentum with parameters $(z_1,z_0)$, as the expressions in terms of Poisson brackets in Eq.~\eqref{eq:RLpoint} indicates. Translations correspond then to constant displacements in the Lagrange multipliers, and translated observers with respect to the local one do not see the worldlines meet at a single space-time point.

Note that the notion of relative locality allows one to \emph{define} an observer with the property of seeing a given high-energy process (that is, one which includes corrections of order $\mathcal{E}/\Lambda$ or $\mathcal{P}/\Lambda$ in the notation of Eq.~\eqref{eq:RLint}) as a local event. This property also defines the origin of space-time coordinates for this observer. The approach that we have followed in Sec.~\ref{sec:td} for calculating the time delay of a high-energy photon is consistent with this idea, since we introduced two observers, $\mathcal{O}$ and $\mathcal{O'}$, who establish their respective origin of space-time coordinates using specific high-energy interactions that they see as local: the detection and the emission of a high-energy photon, respectively.

Since the procedure presented in Sec.~\ref{sec:td} allowed us to relate the space-time coordinates of the worldline of the particle whose detection and emission defines the observers $\mathcal{O}$ and $\mathcal{O'}$, we may ask how observer $\mathcal{O}$ would `see' the process of emission of this particle, which must be nonlocal for this observer, or how $\mathcal{O'}$ would see the process of its detection, that is, which coordinates they would associate to the starting or ending points of worldlines of particles participating in these interactions. 

If we would know the values of $(z_1,z_0)$ in the description by observer $\mathcal{O}$ of the interaction that produces the high-energy photon, one could use Eq.~\eqref{eq:RLpoint} to obtain the space-time coordinates that this observer would assign to the end or start of the worldlines of all the particles participating in the interaction, including the starting point of the trajectory of the high-energy photon. Alternatively, one can determine these last coordinates from Eq.~\eqref{eq:coordtransf} with $x'=t'=0$,
\begin{align}
 x\,&=\, \epsilon_1 \{P,x\} -\epsilon_0 \{E,x\} \,=\, -L \left[ 1+ \frac{(-2a_2+a_3)}{\Lambda}\Pi \right] \,, \notag\\
 t\,&=\, \epsilon_1 \{P,t\} -\epsilon_0 \{E,t\} \,=\, -L \left[ 1 + \frac{(-2a_2+a_3)}{\Lambda} \Pi + \frac{2(a_4+a_5-a_6)}{3\Lambda}\Pi \right]\,. 
\end{align}
The compatibility of the two expressions for the coordinates $(x,t)$ that observer $\mathcal{O}$ assigns to the starting point of the worldline of the high-energy photon allows us to determine the values of the parameters $(z_1,z_0)$ in the description of the emission by observer $\mathcal{O}$. One finds
\begin{align}
z_1 \,&=\, - L \,\left[1 + \frac{(-2 a_2 + a_3)}{\Lambda} \Pi + \left(1 + \frac{\partial\mathcal{E}}{\partial\Pi} - \frac{\partial\mathcal{P}}{\partial\Pi}\right)\right]\,, \notag\\
 z_0 \,&=\, - L \,\left[1 + \frac{(-2 a_2 + a_3)}{\Lambda} \Pi + \frac{2(a_4+a_5-a_6)}{\Lambda} \Pi + \left(1 - \frac{\partial\mathcal{E}}{\partial\Omega} + \frac{\partial\mathcal{P}}{\partial\Omega}\right)\right]\,.  
\end{align}
One has corrections to the result in SR, $z_1=z_0=-L$, due to the deformation of the generators  $(E,P)$ of translations (terms proportional to $(-2a_2+a_3)$),  to the deformation of the boost generator  (term proportional to $(a_4+a_5-a_6)$), and to the modification of the composition of energy and momentum (terms depending on the derivatives of the total energy ($\mathcal{E}$) and total momentum ($\mathcal{P}$) with respect to the energy ($\Omega$) and momentum ($\Pi$) of the high-energy photon).

The method used to determine the parameters $(z_1,z_0)$ in the description of the emission by observer $\mathcal{O}$ can also be used to determine the parameters $(z'_1,z'_0)$, in the description of the detection by observer $\mathcal{O'}$,  reproducing the space-time coordinates $(x',t')$ of the end point of the trajectory of the high-energy photon.

It is remarkable that the derivation of the time delay of a high-energy photon emitted by a source that we got from a model describing its propagation from the source to the detector is consistent with the nonlocality obtained with the action formulation of either its emission or its detection, which incorporates the modified kinematics (modified dispersion relation and modified composition law) of the corresponding interaction. The description of each process by one of the two observers is very complicated and depends on the four-momenta of all the particles involved in the interaction, but the result of the time delay itself is independent of them, and depends only on the four-momentum of the high-energy photon.

Note that we did not try to build a model of multiple interactions, which is not as well-defined as the action formulation of one interaction~\eqref{eq:action}, and which poses important difficulties, such as the spectator problem~\cite{Gubitosi:2019ymi}, to try to obtain an expression for the time delay in DSR. In such a model we would not have been able to treat independently the two interactions (emission and detection of the high energy photon)  and we would have found an obstruction to reproduce the trajectory of the photon, used  in the calculation of the time delay.

\section{Conclusions}
\label{sec:conclusions}

In the present work we have analyzed in detail the determination of time delays in doubly special relativity theories in terms of the parameters that determine the deformation of special relativity. We have clarified long-standing questions related to this problem: first, the independence of the result of an observable, the time delay, of the energy-momentum variables one is free to choose; second, the consistency of the calculation of the time delay with the relative locality of the interactions, leading to a result which is independent of the details of the emission or detection of the high-energy photon; and third, the existence of DSR models without time delays.

The fact that observables in DSR should be independent of the choice of energy-momentum variables seems reasonable and is normally advocated in the literature~\cite{Bevilacqua:2022fnc}, but its physical interpretation was not clear because a change of energy and momentum was usually related to a change of bases in a Hopf algebra, which are assigned to different DSR models~\cite{KowalskiGlikman:2002jr,KowalskiGlikman:2002we}. However, already in Sec.~\ref{sec:rds} we have been very careful to distinguish between the physical energy and momentum, $(\Omega,\Pi)$, which are variables in a canonical phase space, and the generators of translations in spacetime, $E, P$, which are the operators that appear in the deformed Poincaré algebra. We showed in Sec.~\ref{sec:choice} that the result for time delays is indeed independent from the choice of phase-space variables.

Using a canonical phase space as the physical spacetime and energy-momentum is indeed consistent with the action formulation of relative locality that we reviewed in Sec.~\ref{sec:RL}. This formalism gives the space-time coordinates of ending and starting points of worldlines of particles participating in an interaction, as a function of the momenta of all those particles, the total momentum which is conserved in the interaction, and the Lagrange multipliers that a given observer assigns to it. A consequence of relative locality is that each (high-energy) process is only seen as local by a single observer, who assigns zero Lagrange multipliers, corresponding to the origin of space-time coordinates, to the interaction. The procedure of calculation of time delays carried out in Sec.~\ref{sec:td} was indeed compatible with this formulation of relative locality. Remarkably, we obtained a formula for time delays in DSR which depends on the four-momentum of the high-energy photon but is independent of the details of its emission and detection. We also showed in Sec.~\ref{sec:noncommutative} that considering the propagation and detection of the photon in a noncommutative spacetime is an equivalent description up to order $(1/\Lambda)$, that leads to same result for the time delay. 

An important observation is that the nonstandard space-time translations generated by $E$ and $P$, which was the starting point in Sec.~\ref{sec:rds}, have nothing to do with implementing the relative locality of interactions, but with the choice of phase-space variables. In fact, there exists a choice of these variables (when $\Omega=E$, $\Pi=P$) where space-time translations are undeformed in the one-particle system. The translational invariance of DSR is indeed different from that of SR because in DSR, the total momentum which is conserved is not a sum of momenta, but a nonlinear combination of them.  Translations are not, then, constant displacements in the coordinates of the different particles. Deformed translation generators in the one-particle system do not correspond to a translation between observers in the DSR formulation of an interaction, which has the total energy and momentum as the generators of the transformations, and not the functions of the energy and momentum of a single particle $E$ and $P$. 

Finally, we have seen that the expression for the time delay has a sum of two contributions, one due to the momentum dependence of the velocity of propagation of a massless particle, and a second one due to the nontrivial implementation of translational invariance. We have shown that it is possible to have a DSR theory without time delays when these two contributions cancel each other. From the perspective of the representation of deformed Poincaré transformations in terms of the canonical phase-space coordinates, the absence of such an effect requires the condition \eqref{no-time-delay} to be satisfied. However, one can also see the condition in a coordinate-free form, as a condition over the deformed algebra of the generators, Eq~\eqref{no-time-delay-2}. As a consequence, we find the possibility to identify the existence of time delays directly from the deformation of the Casimir. It is also relevant to note that one can have an absence of time delays, and still, a photon dispersion relation (which, let us remark once again, must be seen as a relation between $\Omega$ and $\Pi$, and not between $E$ and $P$), which is different from the one of special relativity.

The absence of time delays in a DSR theory does not contradict the existence of a modified composition law (which we related to the parameters of the relativistic deformation in Sec.~\ref{sec:MCL}), nor the relative locality of the interactions. As commented in the introduction, in the case of DSR there are cancellations of corrections due to the modification of the dispersion relation and the modification of the composition of momenta in the kinematics of processes at energies below the energy scale of the deformation. As a consequence, in contrast to the case of a Lorentz invariance violation, there are no observable effects in the high-energy interactions of particles which are relevant in high-energy astroparticle physics when the scale of the deformation is of the order of the Planck energy scale. This rises the question of which is the most restrictive present bound on the scale of a DSR without time delays and what could be the first observable effect of such a deformation of SR. First steps to answer this question have already started~\cite{Carmona:2021lxr}, but it certainly deserves further investigation. 

\section*{Acknowledgments}
This work is supported by Spanish grants PGC2018-095328-B-I00, funded by MCIN/AEI/10.13039/501100011033
and by ERDF A way of making Europe,  and DGA-FSE grant 2020-E21-17R. JJR acknowledges support from the Unión Europea-NextGenerationEU (``Ayudas Margarita Salas para la formación de jóvenes doctores'').  The work of MAR is supported by MICIU/AEI/FSE (FPI grant PRE2019-089024). This work has been partially supported by Agencia Estatal de Investigaci\'on (Spain)  under grant  PID2019-106802GB-I00/AEI/10.13039/501100011033.  The authors would like to acknowledge the contribution of the COST Action CA18108 ``Quantum gravity phenomenology in the multi-messenger approach''.

\bibliography{QuGraPhenoBib}

\begin{thebibliography}{62}%
\makeatletter
\providecommand \@ifxundefined [1]{%
 \@ifx{#1\undefined}
}%
\providecommand \@ifnum [1]{%
 \ifnum #1\expandafter \@firstoftwo
 \else \expandafter \@secondoftwo
 \fi
}%
\providecommand \@ifx [1]{%
 \ifx #1\expandafter \@firstoftwo
 \else \expandafter \@secondoftwo
 \fi
}%
\providecommand \natexlab [1]{#1}%
\providecommand \enquote  [1]{``#1''}%
\providecommand \bibnamefont  [1]{#1}%
\providecommand \bibfnamefont [1]{#1}%
\providecommand \citenamefont [1]{#1}%
\providecommand \href@noop [0]{\@secondoftwo}%
\providecommand \href [0]{\begingroup \@sanitize@url \@href}%
\providecommand \@href[1]{\@@startlink{#1}\@@href}%
\providecommand \@@href[1]{\endgroup#1\@@endlink}%
\providecommand \@sanitize@url [0]{\catcode `\\12\catcode `\$12\catcode
  `\&12\catcode `\#12\catcode `\^12\catcode `\_12\catcode `\%12\relax}%
\providecommand \@@startlink[1]{}%
\providecommand \@@endlink[0]{}%
\providecommand \url  [0]{\begingroup\@sanitize@url \@url }%
\providecommand \@url [1]{\endgroup\@href {#1}{\urlprefix }}%
\providecommand \urlprefix  [0]{URL }%
\providecommand \Eprint [0]{\href }%
\providecommand \doibase [0]{http://dx.doi.org/}%
\providecommand \selectlanguage [0]{\@gobble}%
\providecommand \bibinfo  [0]{\@secondoftwo}%
\providecommand \bibfield  [0]{\@secondoftwo}%
\providecommand \translation [1]{[#1]}%
\providecommand \BibitemOpen [0]{}%
\providecommand \bibitemStop [0]{}%
\providecommand \bibitemNoStop [0]{.\EOS\space}%
\providecommand \EOS [0]{\spacefactor3000\relax}%
\providecommand \BibitemShut  [1]{\csname bibitem#1\endcsname}%
\let\auto@bib@innerbib\@empty
\bibitem [{\citenamefont {Mukhi}(2011)}]{Mukhi:2011zz}%
  \BibitemOpen
  \bibfield  {author} {\bibinfo {author} {\bibfnamefont {S.}~\bibnamefont
  {Mukhi}},\ }\href {\doibase 10.1088/0264-9381/28/15/153001} {\bibfield
  {journal} {\bibinfo  {journal} {Class. Quant. Grav.}\ }\textbf {\bibinfo
  {volume} {28}},\ \bibinfo {pages} {153001} (\bibinfo {year} {2011})},\
  \Eprint {http://arxiv.org/abs/1110.2569} {arXiv:1110.2569 [physics.pop-ph]}
  \BibitemShut {NoStop}%
\bibitem [{\citenamefont {Aharony}(2000)}]{Aharony:1999ks}%
  \BibitemOpen
  \bibfield  {author} {\bibinfo {author} {\bibfnamefont {O.}~\bibnamefont
  {Aharony}},\ }\bibfield  {booktitle} {\emph {\bibinfo {booktitle} {{Strings
  '99. Proceedings, Conference, Potsdam, Germany, July 19-24, 1999}}},\ }\href
  {\doibase 10.1088/0264-9381/17/5/302} {\bibfield  {journal} {\bibinfo
  {journal} {Class. Quant. Grav.}\ }\textbf {\bibinfo {volume} {17}},\ \bibinfo
  {pages} {929} (\bibinfo {year} {2000})},\ \Eprint
  {http://arxiv.org/abs/hep-th/9911147} {arXiv:hep-th/9911147 [hep-th]}
  \BibitemShut {NoStop}%
\bibitem [{\citenamefont {Dienes}(1997)}]{Dienes:1996du}%
  \BibitemOpen
  \bibfield  {author} {\bibinfo {author} {\bibfnamefont {K.~R.}\ \bibnamefont
  {Dienes}},\ }\bibfield  {booktitle} {\emph {\bibinfo {booktitle} {{Institute
  for Theoretical Physics Conference on Unification: From the Weak Scale to the
  Planck Scale Santa Barbara, California, October 23-27, 1995}}},\ }\href
  {\doibase 10.1016/S0370-1573(97)00009-4} {\bibfield  {journal} {\bibinfo
  {journal} {Phys. Rept.}\ }\textbf {\bibinfo {volume} {287}},\ \bibinfo
  {pages} {447} (\bibinfo {year} {1997})},\ \Eprint
  {http://arxiv.org/abs/hep-th/9602045} {arXiv:hep-th/9602045 [hep-th]}
  \BibitemShut {NoStop}%
\bibitem [{\citenamefont {Sahlmann}(2010)}]{Sahlmann:2010zf}%
  \BibitemOpen
  \bibfield  {author} {\bibinfo {author} {\bibfnamefont {H.}~\bibnamefont
  {Sahlmann}},\ }in\ \href
  {https://inspirehep.net/record/843661/files/arXiv:1001.4188.pdf} {\emph
  {\bibinfo {booktitle} {{Proceedings, Foundations of Space and Time:
  Reflections on Quantum Gravity: Cape Town, South Africa}}}}\ (\bibinfo {year}
  {2010})\ pp.\ \bibinfo {pages} {185--210},\ \Eprint
  {http://arxiv.org/abs/1001.4188} {arXiv:1001.4188 [gr-qc]} \BibitemShut
  {NoStop}%
\bibitem [{\citenamefont {Dupuis}\ \emph {et~al.}(2012)\citenamefont {Dupuis},
  \citenamefont {Ryan},\ and\ \citenamefont {Speziale}}]{Dupuis:2012yw}%
  \BibitemOpen
  \bibfield  {author} {\bibinfo {author} {\bibfnamefont {M.}~\bibnamefont
  {Dupuis}}, \bibinfo {author} {\bibfnamefont {J.~P.}\ \bibnamefont {Ryan}}, \
  and\ \bibinfo {author} {\bibfnamefont {S.}~\bibnamefont {Speziale}},\ }\href
  {\doibase 10.3842/SIGMA.2012.052} {\bibfield  {journal} {\bibinfo  {journal}
  {SIGMA}\ }\textbf {\bibinfo {volume} {8}},\ \bibinfo {pages} {052} (\bibinfo
  {year} {2012})},\ \Eprint {http://arxiv.org/abs/1204.5394} {arXiv:1204.5394
  [gr-qc]} \BibitemShut {NoStop}%
\bibitem [{\citenamefont {Loll}(2019)}]{Loll_2019}%
  \BibitemOpen
  \bibfield  {author} {\bibinfo {author} {\bibfnamefont {R.}~\bibnamefont
  {Loll}},\ }\href {\doibase 10.1088/1361-6382/ab57c7} {\bibfield  {journal}
  {\bibinfo  {journal} {Classical and Quantum Gravity}\ }\textbf {\bibinfo
  {volume} {37}},\ \bibinfo {pages} {013002} (\bibinfo {year}
  {2019})}\BibitemShut {NoStop}%
\bibitem [{\citenamefont {Wallden}(2013)}]{Wallden:2013kka}%
  \BibitemOpen
  \bibfield  {author} {\bibinfo {author} {\bibfnamefont {P.}~\bibnamefont
  {Wallden}},\ }\bibfield  {booktitle} {\emph {\bibinfo {booktitle}
  {{Proceedings, 15th Conference on Recent Developments in Gravity (NEB 15):
  Chania, Crete, Greece, June 20-23, 2012}}},\ }\href {\doibase
  10.1088/1742-6596/453/1/012023} {\bibfield  {journal} {\bibinfo  {journal}
  {J. Phys. Conf. Ser.}\ }\textbf {\bibinfo {volume} {453}},\ \bibinfo {pages}
  {012023} (\bibinfo {year} {2013})}\BibitemShut {NoStop}%
\bibitem [{\citenamefont {Wallden}(2010)}]{Wallden:2010sh}%
  \BibitemOpen
  \bibfield  {author} {\bibinfo {author} {\bibfnamefont {P.}~\bibnamefont
  {Wallden}},\ }\bibfield  {booktitle} {\emph {\bibinfo {booktitle} {{Classical
  and quantum gravity. Proceedings, 1st Mediterranean Conference, MCCQG 2009,
  Kolymbari, Crete, Greece, September 14-18, 2009}}},\ }\href {\doibase
  10.1088/1742-6596/222/1/012053} {\bibfield  {journal} {\bibinfo  {journal}
  {J. Phys. Conf. Ser.}\ }\textbf {\bibinfo {volume} {222}},\ \bibinfo {pages}
  {012053} (\bibinfo {year} {2010})},\ \Eprint {http://arxiv.org/abs/1001.4041}
  {arXiv:1001.4041 [gr-qc]} \BibitemShut {NoStop}%
\bibitem [{\citenamefont {Henson}(2009)}]{Henson:2006kf}%
  \BibitemOpen
  \bibfield  {author} {\bibinfo {author} {\bibfnamefont {J.}~\bibnamefont
  {Henson}},\ }in\ \href@noop {} {\emph {\bibinfo {booktitle} {Approaches to
  Quantum Gravity: Toward a New Understanding of Space, Time and Matter}}},\
  \bibinfo {editor} {edited by\ \bibinfo {editor} {\bibfnamefont
  {D.}~\bibnamefont {Oriti}}}\ (\bibinfo  {publisher} {Cambridge University
  Press},\ \bibinfo {year} {2009})\ pp.\ \bibinfo {pages} {393--413},\ \Eprint
  {http://arxiv.org/abs/gr-qc/0601121} {arXiv:gr-qc/0601121 [gr-qc]}
  \BibitemShut {NoStop}%
\bibitem [{\citenamefont {Gross}\ and\ \citenamefont
  {Mende}(1988)}]{Gross:1987ar}%
  \BibitemOpen
  \bibfield  {author} {\bibinfo {author} {\bibfnamefont {D.~J.}\ \bibnamefont
  {Gross}}\ and\ \bibinfo {author} {\bibfnamefont {P.~F.}\ \bibnamefont
  {Mende}},\ }\href {\doibase 10.1016/0550-3213(88)90390-2} {\bibfield
  {journal} {\bibinfo  {journal} {Nucl. Phys.}\ }\textbf {\bibinfo {volume}
  {B303}},\ \bibinfo {pages} {407} (\bibinfo {year} {1988})}\BibitemShut
  {NoStop}%
\bibitem [{\citenamefont {Amati}\ \emph {et~al.}(1989)\citenamefont {Amati},
  \citenamefont {Ciafaloni},\ and\ \citenamefont {Veneziano}}]{Amati:1988tn}%
  \BibitemOpen
  \bibfield  {author} {\bibinfo {author} {\bibfnamefont {D.}~\bibnamefont
  {Amati}}, \bibinfo {author} {\bibfnamefont {M.}~\bibnamefont {Ciafaloni}}, \
  and\ \bibinfo {author} {\bibfnamefont {G.}~\bibnamefont {Veneziano}},\ }\href
  {\doibase 10.1016/0370-2693(89)91366-X} {\bibfield  {journal} {\bibinfo
  {journal} {Phys. Lett.}\ }\textbf {\bibinfo {volume} {B216}},\ \bibinfo
  {pages} {41} (\bibinfo {year} {1989})}\BibitemShut {NoStop}%
\bibitem [{\citenamefont {Garay}(1995)}]{Garay1995}%
  \BibitemOpen
  \bibfield  {author} {\bibinfo {author} {\bibfnamefont {L.~J.}\ \bibnamefont
  {Garay}},\ }\href {\doibase 10.1142/S0217751X95000085} {\bibfield  {journal}
  {\bibinfo  {journal} {Int. J. Mod. Phys.}\ }\textbf {\bibinfo {volume}
  {A10}},\ \bibinfo {pages} {145} (\bibinfo {year} {1995})},\ \Eprint
  {http://arxiv.org/abs/gr-qc/9403008} {arXiv:gr-qc/9403008 [gr-qc]}
  \BibitemShut {NoStop}%
\bibitem [{\citenamefont {Addazi}\ \emph {et~al.}(2022)\citenamefont {Addazi}
  \emph {et~al.}}]{Addazi:2021xuf}%
  \BibitemOpen
  \bibfield  {author} {\bibinfo {author} {\bibfnamefont {A.}~\bibnamefont
  {Addazi}} \emph {et~al.},\ }\href {\doibase 10.1016/j.ppnp.2022.103948}
  {\bibfield  {journal} {\bibinfo  {journal} {Prog. Part. Nucl. Phys.}\
  }\textbf {\bibinfo {volume} {125}},\ \bibinfo {pages} {103948} (\bibinfo
  {year} {2022})},\ \Eprint {http://arxiv.org/abs/2111.05659} {arXiv:2111.05659
  [hep-ph]} \BibitemShut {NoStop}%
\bibitem [{\citenamefont {Colladay}\ and\ \citenamefont
  {Kostelecky}(1998)}]{Colladay:1998fq}%
  \BibitemOpen
  \bibfield  {author} {\bibinfo {author} {\bibfnamefont {D.}~\bibnamefont
  {Colladay}}\ and\ \bibinfo {author} {\bibfnamefont {V.~A.}\ \bibnamefont
  {Kostelecky}},\ }\href {\doibase 10.1103/PhysRevD.58.116002} {\bibfield
  {journal} {\bibinfo  {journal} {Phys. Rev.}\ }\textbf {\bibinfo {volume}
  {D58}},\ \bibinfo {pages} {116002} (\bibinfo {year} {1998})},\ \Eprint
  {http://arxiv.org/abs/hep-ph/9809521} {arXiv:hep-ph/9809521 [hep-ph]}
  \BibitemShut {NoStop}%
\bibitem [{\citenamefont {Kostelecky}\ and\ \citenamefont
  {Russell}(2011)}]{Kostelecky:2008ts}%
  \BibitemOpen
  \bibfield  {author} {\bibinfo {author} {\bibfnamefont {V.~A.}\ \bibnamefont
  {Kostelecky}}\ and\ \bibinfo {author} {\bibfnamefont {N.}~\bibnamefont
  {Russell}},\ }\href {\doibase 10.1103/RevModPhys.83.11} {\bibfield  {journal}
  {\bibinfo  {journal} {Rev. Mod. Phys.}\ }\textbf {\bibinfo {volume} {83}},\
  \bibinfo {pages} {11} (\bibinfo {year} {2011})},\ \Eprint
  {http://arxiv.org/abs/0801.0287} {arXiv:0801.0287 [hep-ph]} \BibitemShut
  {NoStop}%
\bibitem [{\citenamefont {Amelino-Camelia}(2013)}]{AmelinoCamelia:2008qg}%
  \BibitemOpen
  \bibfield  {author} {\bibinfo {author} {\bibfnamefont {G.}~\bibnamefont
  {Amelino-Camelia}},\ }\href {\doibase 10.12942/lrr-2013-5} {\bibfield
  {journal} {\bibinfo  {journal} {Living Rev.Rel.}\ }\textbf {\bibinfo {volume}
  {16}},\ \bibinfo {pages} {5} (\bibinfo {year} {2013})},\ \Eprint
  {http://arxiv.org/abs/0806.0339} {arXiv:0806.0339 [gr-qc]} \BibitemShut
  {NoStop}%
\bibitem [{\citenamefont {Majid}\ and\ \citenamefont
  {Ruegg}(1994)}]{Majid1994}%
  \BibitemOpen
  \bibfield  {author} {\bibinfo {author} {\bibfnamefont {S.}~\bibnamefont
  {Majid}}\ and\ \bibinfo {author} {\bibfnamefont {H.}~\bibnamefont {Ruegg}},\
  }\href {\doibase 10.1016/0370-2693(94)90699-8} {\bibfield  {journal}
  {\bibinfo  {journal} {Phys. Lett.}\ }\textbf {\bibinfo {volume} {B334}},\
  \bibinfo {pages} {348} (\bibinfo {year} {1994})},\ \Eprint
  {http://arxiv.org/abs/hep-th/9405107} {arXiv:hep-th/9405107 [hep-th]}
  \BibitemShut {NoStop}%
\bibitem [{\citenamefont {Lukierski}\ \emph {et~al.}(1991)\citenamefont
  {Lukierski}, \citenamefont {Ruegg}, \citenamefont {Nowicki},\ and\
  \citenamefont {Tolstoi}}]{Lukierski:1991pn}%
  \BibitemOpen
  \bibfield  {author} {\bibinfo {author} {\bibfnamefont {J.}~\bibnamefont
  {Lukierski}}, \bibinfo {author} {\bibfnamefont {H.}~\bibnamefont {Ruegg}},
  \bibinfo {author} {\bibfnamefont {A.}~\bibnamefont {Nowicki}}, \ and\
  \bibinfo {author} {\bibfnamefont {V.~N.}\ \bibnamefont {Tolstoi}},\ }\href
  {\doibase 10.1016/0370-2693(91)90358-W} {\bibfield  {journal} {\bibinfo
  {journal} {Phys. Lett.}\ }\textbf {\bibinfo {volume} {B264}},\ \bibinfo
  {pages} {331} (\bibinfo {year} {1991})}\BibitemShut {NoStop}%
\bibitem [{\citenamefont {Lukierski}\ \emph {et~al.}(1993)\citenamefont
  {Lukierski}, \citenamefont {Ruegg},\ and\ \citenamefont
  {Ruhl}}]{Lukierski:1993df}%
  \BibitemOpen
  \bibfield  {author} {\bibinfo {author} {\bibfnamefont {J.}~\bibnamefont
  {Lukierski}}, \bibinfo {author} {\bibfnamefont {H.}~\bibnamefont {Ruegg}}, \
  and\ \bibinfo {author} {\bibfnamefont {W.}~\bibnamefont {Ruhl}},\ }\href
  {\doibase 10.1016/0370-2693(93)90004-2} {\bibfield  {journal} {\bibinfo
  {journal} {Phys. Lett.}\ }\textbf {\bibinfo {volume} {B313}},\ \bibinfo
  {pages} {357} (\bibinfo {year} {1993})}\BibitemShut {NoStop}%
\bibitem [{\citenamefont {Lukierski}\ \emph {et~al.}(1992)\citenamefont
  {Lukierski}, \citenamefont {Nowicki},\ and\ \citenamefont
  {Ruegg}}]{Lukierski:1992dt}%
  \BibitemOpen
  \bibfield  {author} {\bibinfo {author} {\bibfnamefont {J.}~\bibnamefont
  {Lukierski}}, \bibinfo {author} {\bibfnamefont {A.}~\bibnamefont {Nowicki}},
  \ and\ \bibinfo {author} {\bibfnamefont {H.}~\bibnamefont {Ruegg}},\ }\href
  {\doibase 10.1016/0370-2693(92)90894-A} {\bibfield  {journal} {\bibinfo
  {journal} {Phys. Lett.}\ }\textbf {\bibinfo {volume} {B293}},\ \bibinfo
  {pages} {344} (\bibinfo {year} {1992})}\BibitemShut {NoStop}%
\bibitem [{\citenamefont {Lukierski}\ and\ \citenamefont
  {Nowicki}(2003)}]{Lukierski:2002df}%
  \BibitemOpen
  \bibfield  {author} {\bibinfo {author} {\bibfnamefont {J.}~\bibnamefont
  {Lukierski}}\ and\ \bibinfo {author} {\bibfnamefont {A.}~\bibnamefont
  {Nowicki}},\ }\href {\doibase 10.1142/S0217751X03013600} {\bibfield
  {journal} {\bibinfo  {journal} {Int. J. Mod. Phys.}\ }\textbf {\bibinfo
  {volume} {A18}},\ \bibinfo {pages} {7} (\bibinfo {year} {2003})},\ \Eprint
  {http://arxiv.org/abs/hep-th/0203065} {arXiv:hep-th/0203065 [hep-th]}
  \BibitemShut {NoStop}%
\bibitem [{\citenamefont {Carmona}\ \emph
  {et~al.}(2019{\natexlab{a}})\citenamefont {Carmona}, \citenamefont
  {Cortés},\ and\ \citenamefont {Relancio}}]{Carmona:2019fwf}%
  \BibitemOpen
  \bibfield  {author} {\bibinfo {author} {\bibfnamefont {J.~M.}\ \bibnamefont
  {Carmona}}, \bibinfo {author} {\bibfnamefont {J.~L.}\ \bibnamefont
  {Cortés}}, \ and\ \bibinfo {author} {\bibfnamefont {J.~J.}\ \bibnamefont
  {Relancio}},\ }\href {\doibase 10.1103/PhysRevD.100.104031} {\bibfield
  {journal} {\bibinfo  {journal} {Phys. Rev.}\ }\textbf {\bibinfo {volume}
  {D100}},\ \bibinfo {pages} {104031} (\bibinfo {year} {2019}{\natexlab{a}})},\
  \Eprint {http://arxiv.org/abs/1907.12298} {arXiv:1907.12298 [hep-th]}
  \BibitemShut {NoStop}%
\bibitem [{\citenamefont {Kowalski-Glikman}\ and\ \citenamefont
  {Nowak}(2003)}]{KowalskiGlikman:2002jr}%
  \BibitemOpen
  \bibfield  {author} {\bibinfo {author} {\bibfnamefont {J.}~\bibnamefont
  {Kowalski-Glikman}}\ and\ \bibinfo {author} {\bibfnamefont {S.}~\bibnamefont
  {Nowak}},\ }\href {\doibase 10.1142/S0218271803003050} {\bibfield  {journal}
  {\bibinfo  {journal} {Int. J. Mod. Phys.}\ }\textbf {\bibinfo {volume}
  {D12}},\ \bibinfo {pages} {299} (\bibinfo {year} {2003})},\ \Eprint
  {http://arxiv.org/abs/hep-th/0204245} {arXiv:hep-th/0204245 [hep-th]}
  \BibitemShut {NoStop}%
\bibitem [{\citenamefont {Borowiec}\ and\ \citenamefont
  {Pachol}(2010)}]{Borowiec2010}%
  \BibitemOpen
  \bibfield  {author} {\bibinfo {author} {\bibfnamefont {A.}~\bibnamefont
  {Borowiec}}\ and\ \bibinfo {author} {\bibfnamefont {A.}~\bibnamefont
  {Pachol}},\ }\href {\doibase 10.1088/1751-8113/43/4/045203} {\bibfield
  {journal} {\bibinfo  {journal} {J. Phys.}\ }\textbf {\bibinfo {volume}
  {A43}},\ \bibinfo {pages} {045203} (\bibinfo {year} {2010})},\ \Eprint
  {http://arxiv.org/abs/0903.5251} {arXiv:0903.5251 [hep-th]} \BibitemShut
  {NoStop}%
\bibitem [{\citenamefont {Mattingly}(2005)}]{Mattingly:2005re}%
  \BibitemOpen
  \bibfield  {author} {\bibinfo {author} {\bibfnamefont {D.}~\bibnamefont
  {Mattingly}},\ }\href@noop {} {\bibfield  {journal} {\bibinfo  {journal}
  {Living Rev.Rel.}\ }\textbf {\bibinfo {volume} {8}},\ \bibinfo {pages} {5}
  (\bibinfo {year} {2005})},\ \Eprint {http://arxiv.org/abs/gr-qc/0502097}
  {arXiv:gr-qc/0502097 [gr-qc]} \BibitemShut {NoStop}%
\bibitem [{\citenamefont {Liberati}(2013)}]{Liberati:2013xla}%
  \BibitemOpen
  \bibfield  {author} {\bibinfo {author} {\bibfnamefont {S.}~\bibnamefont
  {Liberati}},\ }\href {\doibase 10.1088/0264-9381/30/13/133001} {\bibfield
  {journal} {\bibinfo  {journal} {Class. Quant. Grav.}\ }\textbf {\bibinfo
  {volume} {30}},\ \bibinfo {pages} {133001} (\bibinfo {year} {2013})},\
  \Eprint {http://arxiv.org/abs/1304.5795} {arXiv:1304.5795 [gr-qc]}
  \BibitemShut {NoStop}%
\bibitem [{\citenamefont {Albalate}\ \emph {et~al.}(2018)\citenamefont
  {Albalate}, \citenamefont {Carmona}, \citenamefont {Cort\'es},\ and\
  \citenamefont {Relancio}}]{Albalate:2018kcf}%
  \BibitemOpen
  \bibfield  {author} {\bibinfo {author} {\bibfnamefont {G.}~\bibnamefont
  {Albalate}}, \bibinfo {author} {\bibfnamefont {J.~M.}\ \bibnamefont
  {Carmona}}, \bibinfo {author} {\bibfnamefont {J.~L.}\ \bibnamefont
  {Cort\'es}}, \ and\ \bibinfo {author} {\bibfnamefont {J.~J.}\ \bibnamefont
  {Relancio}},\ }\href {\doibase 10.3390/sym10100432} {\bibfield  {journal}
  {\bibinfo  {journal} {Symmetry}\ }\textbf {\bibinfo {volume} {10}},\ \bibinfo
  {pages} {432} (\bibinfo {year} {2018})},\ \Eprint
  {http://arxiv.org/abs/1809.08167} {arXiv:1809.08167 [hep-ph]} \BibitemShut
  {NoStop}%
\bibitem [{\citenamefont {Carmona}\ \emph
  {et~al.}(2020{\natexlab{a}})\citenamefont {Carmona}, \citenamefont
  {Cort\'es}, \citenamefont {Pereira},\ and\ \citenamefont
  {Relancio}}]{Carmona:2020whi}%
  \BibitemOpen
  \bibfield  {author} {\bibinfo {author} {\bibfnamefont {J.~M.}\ \bibnamefont
  {Carmona}}, \bibinfo {author} {\bibfnamefont {J.~L.}\ \bibnamefont
  {Cort\'es}}, \bibinfo {author} {\bibfnamefont {L.}~\bibnamefont {Pereira}}, \
  and\ \bibinfo {author} {\bibfnamefont {J.~J.}\ \bibnamefont {Relancio}},\
  }\href {\doibase 10.3390/sym12081298} {\bibfield  {journal} {\bibinfo
  {journal} {Symmetry}\ }\textbf {\bibinfo {volume} {12}},\ \bibinfo {pages}
  {1298} (\bibinfo {year} {2020}{\natexlab{a}})},\ \Eprint
  {http://arxiv.org/abs/2008.10251} {arXiv:2008.10251 [hep-ph]} \BibitemShut
  {NoStop}%
\bibitem [{\citenamefont {Relancio}\ and\ \citenamefont
  {Liberati}(2020)}]{Relancio:2020mpa}%
  \BibitemOpen
  \bibfield  {author} {\bibinfo {author} {\bibfnamefont {J.}~\bibnamefont
  {Relancio}}\ and\ \bibinfo {author} {\bibfnamefont {S.}~\bibnamefont
  {Liberati}},\ }\href {\doibase 10.1103/PhysRevD.102.104025} {\bibfield
  {journal} {\bibinfo  {journal} {Phys. Rev. D}\ }\textbf {\bibinfo {volume}
  {102}},\ \bibinfo {pages} {104025} (\bibinfo {year} {2020})},\ \Eprint
  {http://arxiv.org/abs/2008.08317} {arXiv:2008.08317 [gr-qc]} \BibitemShut
  {NoStop}%
\bibitem [{\citenamefont {Carmona}\ \emph {et~al.}(2022)\citenamefont
  {Carmona}, \citenamefont {Cort\'es}, \citenamefont {Relancio}, \citenamefont
  {Reyes},\ and\ \citenamefont {Vincueria}}]{Carmona:2021lxr}%
  \BibitemOpen
  \bibfield  {author} {\bibinfo {author} {\bibfnamefont {J.~M.}\ \bibnamefont
  {Carmona}}, \bibinfo {author} {\bibfnamefont {J.~L.}\ \bibnamefont
  {Cort\'es}}, \bibinfo {author} {\bibfnamefont {J.~J.}\ \bibnamefont
  {Relancio}}, \bibinfo {author} {\bibfnamefont {M.~A.}\ \bibnamefont {Reyes}},
  \ and\ \bibinfo {author} {\bibfnamefont {A.}~\bibnamefont {Vincueria}},\
  }\href {\doibase 10.1140/epjp/s13360-022-02920-3} {\bibfield  {journal}
  {\bibinfo  {journal} {Eur. Phys. J. Plus}\ }\textbf {\bibinfo {volume}
  {137}},\ \bibinfo {pages} {768} (\bibinfo {year} {2022})},\ \Eprint
  {http://arxiv.org/abs/2109.08402} {arXiv:2109.08402 [hep-ph]} \BibitemShut
  {NoStop}%
\bibitem [{\citenamefont {Ellis}\ \emph {et~al.}(2003)\citenamefont {Ellis},
  \citenamefont {Mavromatos}, \citenamefont {Nanopoulos},\ and\ \citenamefont
  {Sakharov}}]{Ellis:2002in}%
  \BibitemOpen
  \bibfield  {author} {\bibinfo {author} {\bibfnamefont {J.~R.}\ \bibnamefont
  {Ellis}}, \bibinfo {author} {\bibfnamefont {N.}~\bibnamefont {Mavromatos}},
  \bibinfo {author} {\bibfnamefont {D.~V.}\ \bibnamefont {Nanopoulos}}, \ and\
  \bibinfo {author} {\bibfnamefont {A.~S.}\ \bibnamefont {Sakharov}},\ }\href
  {\doibase 10.1051/0004-6361:20030263} {\bibfield  {journal} {\bibinfo
  {journal} {Astron. Astrophys.}\ }\textbf {\bibinfo {volume} {402}},\ \bibinfo
  {pages} {409} (\bibinfo {year} {2003})},\ \Eprint
  {http://arxiv.org/abs/astro-ph/0210124} {arXiv:astro-ph/0210124} \BibitemShut
  {NoStop}%
\bibitem [{\citenamefont {Rodriguez~Martinez}\ and\ \citenamefont
  {Piran}(2006)}]{RodriguezMartinez:2006ee}%
  \BibitemOpen
  \bibfield  {author} {\bibinfo {author} {\bibfnamefont {M.}~\bibnamefont
  {Rodriguez~Martinez}}\ and\ \bibinfo {author} {\bibfnamefont
  {T.}~\bibnamefont {Piran}},\ }\href {\doibase 10.1088/1475-7516/2006/04/006}
  {\bibfield  {journal} {\bibinfo  {journal} {JCAP}\ }\textbf {\bibinfo
  {volume} {04}},\ \bibinfo {pages} {006} (\bibinfo {year} {2006})},\ \Eprint
  {http://arxiv.org/abs/astro-ph/0601219} {arXiv:astro-ph/0601219} \BibitemShut
  {NoStop}%
\bibitem [{\citenamefont {Jacob}\ and\ \citenamefont
  {Piran}(2008)}]{Jacob:2008bw}%
  \BibitemOpen
  \bibfield  {author} {\bibinfo {author} {\bibfnamefont {U.}~\bibnamefont
  {Jacob}}\ and\ \bibinfo {author} {\bibfnamefont {T.}~\bibnamefont {Piran}},\
  }\href {\doibase 10.1088/1475-7516/2008/01/031} {\bibfield  {journal}
  {\bibinfo  {journal} {JCAP}\ }\textbf {\bibinfo {volume} {01}},\ \bibinfo
  {pages} {031} (\bibinfo {year} {2008})},\ \Eprint
  {http://arxiv.org/abs/0712.2170} {arXiv:0712.2170 [astro-ph]} \BibitemShut
  {NoStop}%
\bibitem [{\citenamefont {Abramowski}\ \emph {et~al.}(2011)\citenamefont
  {Abramowski} \emph {et~al.}}]{HESS:2011aa}%
  \BibitemOpen
  \bibfield  {author} {\bibinfo {author} {\bibfnamefont {A.}~\bibnamefont
  {Abramowski}} \emph {et~al.} (\bibinfo {collaboration} {HESS}),\ }\href
  {\doibase 10.1016/j.astropartphys.2011.01.007} {\bibfield  {journal}
  {\bibinfo  {journal} {Astropart. Phys.}\ }\textbf {\bibinfo {volume} {34}},\
  \bibinfo {pages} {738} (\bibinfo {year} {2011})},\ \Eprint
  {http://arxiv.org/abs/1101.3650} {arXiv:1101.3650 [astro-ph.HE]} \BibitemShut
  {NoStop}%
\bibitem [{\citenamefont {Abdalla}\ \emph {et~al.}(2019)\citenamefont {Abdalla}
  \emph {et~al.}}]{HESS:2019rhe}%
  \BibitemOpen
  \bibfield  {author} {\bibinfo {author} {\bibfnamefont {H.}~\bibnamefont
  {Abdalla}} \emph {et~al.} (\bibinfo {collaboration} {H.E.S.S.}),\ }\href
  {\doibase 10.3847/1538-4357/aaf1c4} {\bibfield  {journal} {\bibinfo
  {journal} {Astrophys. J.}\ }\textbf {\bibinfo {volume} {870}},\ \bibinfo
  {pages} {93} (\bibinfo {year} {2019})},\ \Eprint
  {http://arxiv.org/abs/1901.05209} {arXiv:1901.05209 [astro-ph.HE]}
  \BibitemShut {NoStop}%
\bibitem [{\citenamefont {Acciari}\ \emph {et~al.}(2020)\citenamefont {Acciari}
  \emph {et~al.}}]{MAGIC:2020egb}%
  \BibitemOpen
  \bibfield  {author} {\bibinfo {author} {\bibfnamefont {V.~A.}\ \bibnamefont
  {Acciari}} \emph {et~al.} (\bibinfo {collaboration} {MAGIC, Armenian
  Consortium: ICRANet-Armenia at NAS RA, A. Alikhanyan National Laboratory,
  Finnish MAGIC Consortium: Finnish Centre of Astronomy with ESO}),\ }\href
  {\doibase 10.1103/PhysRevLett.125.021301} {\bibfield  {journal} {\bibinfo
  {journal} {Phys. Rev. Lett.}\ }\textbf {\bibinfo {volume} {125}},\ \bibinfo
  {pages} {021301} (\bibinfo {year} {2020})},\ \Eprint
  {http://arxiv.org/abs/2001.09728} {arXiv:2001.09728 [astro-ph.HE]}
  \BibitemShut {NoStop}%
\bibitem [{\citenamefont {Du}\ \emph {et~al.}(2021)\citenamefont {Du} \emph
  {et~al.}}]{Du:2020uev}%
  \BibitemOpen
  \bibfield  {author} {\bibinfo {author} {\bibfnamefont {S.-S.}\ \bibnamefont
  {Du}} \emph {et~al.},\ }\href {\doibase 10.3847/1538-4357/abc624} {\bibfield
  {journal} {\bibinfo  {journal} {Astrophys. J.}\ }\textbf {\bibinfo {volume}
  {906}},\ \bibinfo {pages} {8} (\bibinfo {year} {2021})},\ \Eprint
  {http://arxiv.org/abs/2010.16029} {arXiv:2010.16029 [astro-ph.HE]}
  \BibitemShut {NoStop}%
\bibitem [{\citenamefont {Martinez}\ and\ \citenamefont
  {Errando}(2009)}]{Martinez:2008ki}%
  \BibitemOpen
  \bibfield  {author} {\bibinfo {author} {\bibfnamefont {M.}~\bibnamefont
  {Martinez}}\ and\ \bibinfo {author} {\bibfnamefont {M.}~\bibnamefont
  {Errando}},\ }\href {\doibase 10.1016/j.astropartphys.2009.01.005} {\bibfield
   {journal} {\bibinfo  {journal} {Astropart. Phys.}\ }\textbf {\bibinfo
  {volume} {31}},\ \bibinfo {pages} {226} (\bibinfo {year} {2009})},\ \Eprint
  {http://arxiv.org/abs/0803.2120} {arXiv:0803.2120 [astro-ph]} \BibitemShut
  {NoStop}%
\bibitem [{\citenamefont {Vasileiou}\ \emph {et~al.}(2013)\citenamefont
  {Vasileiou}, \citenamefont {Jacholkowska}, \citenamefont {Piron},
  \citenamefont {Bolmont}, \citenamefont {Couturier}, \citenamefont {Granot},
  \citenamefont {Stecker}, \citenamefont {Cohen-Tanugi},\ and\ \citenamefont
  {Longo}}]{Vasileiou:2013vra}%
  \BibitemOpen
  \bibfield  {author} {\bibinfo {author} {\bibfnamefont {V.}~\bibnamefont
  {Vasileiou}}, \bibinfo {author} {\bibfnamefont {A.}~\bibnamefont
  {Jacholkowska}}, \bibinfo {author} {\bibfnamefont {F.}~\bibnamefont {Piron}},
  \bibinfo {author} {\bibfnamefont {J.}~\bibnamefont {Bolmont}}, \bibinfo
  {author} {\bibfnamefont {C.}~\bibnamefont {Couturier}}, \bibinfo {author}
  {\bibfnamefont {J.}~\bibnamefont {Granot}}, \bibinfo {author} {\bibfnamefont
  {F.~W.}\ \bibnamefont {Stecker}}, \bibinfo {author} {\bibfnamefont
  {J.}~\bibnamefont {Cohen-Tanugi}}, \ and\ \bibinfo {author} {\bibfnamefont
  {F.}~\bibnamefont {Longo}},\ }\href {\doibase 10.1103/PhysRevD.87.122001}
  {\bibfield  {journal} {\bibinfo  {journal} {Phys. Rev.}\ }\textbf {\bibinfo
  {volume} {D87}},\ \bibinfo {pages} {122001} (\bibinfo {year} {2013})},\
  \Eprint {http://arxiv.org/abs/1305.3463} {arXiv:1305.3463 [astro-ph.HE]}
  \BibitemShut {NoStop}%
\bibitem [{\citenamefont {Ahnen}\ \emph {et~al.}(2017)\citenamefont {Ahnen}
  \emph {et~al.}}]{MAGIC:2017vah}%
  \BibitemOpen
  \bibfield  {author} {\bibinfo {author} {\bibfnamefont {M.~L.}\ \bibnamefont
  {Ahnen}} \emph {et~al.} (\bibinfo {collaboration} {MAGIC}),\ }\href {\doibase
  10.3847/1538-4365/aa8404} {\bibfield  {journal} {\bibinfo  {journal}
  {Astrophys. J. Suppl.}\ }\textbf {\bibinfo {volume} {232}},\ \bibinfo {pages}
  {9} (\bibinfo {year} {2017})},\ \Eprint {http://arxiv.org/abs/1709.00346}
  {arXiv:1709.00346 [astro-ph.HE]} \BibitemShut {NoStop}%
\bibitem [{\citenamefont {Xu}\ and\ \citenamefont {Ma}(2016)}]{Xu:2016zxi}%
  \BibitemOpen
  \bibfield  {author} {\bibinfo {author} {\bibfnamefont {H.}~\bibnamefont
  {Xu}}\ and\ \bibinfo {author} {\bibfnamefont {B.-Q.}\ \bibnamefont {Ma}},\
  }\href {\doibase 10.1016/j.astropartphys.2016.05.008} {\bibfield  {journal}
  {\bibinfo  {journal} {Astropart. Phys.}\ }\textbf {\bibinfo {volume} {82}},\
  \bibinfo {pages} {72} (\bibinfo {year} {2016})},\ \Eprint
  {http://arxiv.org/abs/1607.03203} {arXiv:1607.03203 [hep-ph]} \BibitemShut
  {NoStop}%
\bibitem [{\citenamefont {Li}\ and\ \citenamefont {Ma}(2021)}]{Li:2021tcw}%
  \BibitemOpen
  \bibfield  {author} {\bibinfo {author} {\bibfnamefont {C.}~\bibnamefont
  {Li}}\ and\ \bibinfo {author} {\bibfnamefont {B.-Q.}\ \bibnamefont {Ma}},\
  }\href {\doibase 10.1103/PhysRevD.104.063012} {\bibfield  {journal} {\bibinfo
   {journal} {Phys. Rev. D}\ }\textbf {\bibinfo {volume} {104}},\ \bibinfo
  {pages} {063012} (\bibinfo {year} {2021})},\ \Eprint
  {http://arxiv.org/abs/2105.07967} {arXiv:2105.07967 [astro-ph.HE]}
  \BibitemShut {NoStop}%
\bibitem [{\citenamefont {Amelino-Camelia}\ \emph
  {et~al.}(2011{\natexlab{a}})\citenamefont {Amelino-Camelia}, \citenamefont
  {Loret},\ and\ \citenamefont {Rosati}}]{Amelino-Camelia:2011ebd}%
  \BibitemOpen
  \bibfield  {author} {\bibinfo {author} {\bibfnamefont {G.}~\bibnamefont
  {Amelino-Camelia}}, \bibinfo {author} {\bibfnamefont {N.}~\bibnamefont
  {Loret}}, \ and\ \bibinfo {author} {\bibfnamefont {G.}~\bibnamefont
  {Rosati}},\ }\href {\doibase 10.1016/j.physletb.2011.04.054} {\bibfield
  {journal} {\bibinfo  {journal} {Phys. Lett. B}\ }\textbf {\bibinfo {volume}
  {700}},\ \bibinfo {pages} {150} (\bibinfo {year} {2011}{\natexlab{a}})},\
  \Eprint {http://arxiv.org/abs/1102.4637} {arXiv:1102.4637 [hep-th]}
  \BibitemShut {NoStop}%
\bibitem [{\citenamefont {Loret}(2014)}]{Loret:2014uia}%
  \BibitemOpen
  \bibfield  {author} {\bibinfo {author} {\bibfnamefont {N.}~\bibnamefont
  {Loret}},\ }\href {\doibase 10.1103/PhysRevD.90.124013} {\bibfield  {journal}
  {\bibinfo  {journal} {Phys. Rev. D}\ }\textbf {\bibinfo {volume} {90}},\
  \bibinfo {pages} {124013} (\bibinfo {year} {2014})},\ \Eprint
  {http://arxiv.org/abs/1404.5093} {arXiv:1404.5093 [hep-th]} \BibitemShut
  {NoStop}%
\bibitem [{\citenamefont {Rosati}\ \emph {et~al.}(2015)\citenamefont {Rosati},
  \citenamefont {Amelino-Camelia}, \citenamefont {Marciano},\ and\
  \citenamefont {Matassa}}]{Rosati:2015pga}%
  \BibitemOpen
  \bibfield  {author} {\bibinfo {author} {\bibfnamefont {G.}~\bibnamefont
  {Rosati}}, \bibinfo {author} {\bibfnamefont {G.}~\bibnamefont
  {Amelino-Camelia}}, \bibinfo {author} {\bibfnamefont {A.}~\bibnamefont
  {Marciano}}, \ and\ \bibinfo {author} {\bibfnamefont {M.}~\bibnamefont
  {Matassa}},\ }\href {\doibase 10.1103/PhysRevD.92.124042} {\bibfield
  {journal} {\bibinfo  {journal} {Phys. Rev.}\ }\textbf {\bibinfo {volume}
  {D92}},\ \bibinfo {pages} {124042} (\bibinfo {year} {2015})},\ \Eprint
  {http://arxiv.org/abs/1507.02056} {arXiv:1507.02056 [hep-th]} \BibitemShut
  {NoStop}%
\bibitem [{\citenamefont {Mignemi}\ and\ \citenamefont
  {Samsarov}(2017)}]{Mignemi:2016ilu}%
  \BibitemOpen
  \bibfield  {author} {\bibinfo {author} {\bibfnamefont {S.}~\bibnamefont
  {Mignemi}}\ and\ \bibinfo {author} {\bibfnamefont {A.}~\bibnamefont
  {Samsarov}},\ }\href {\doibase 10.1016/j.physleta.2017.03.033} {\bibfield
  {journal} {\bibinfo  {journal} {Phys. Lett.}\ }\textbf {\bibinfo {volume}
  {A381}},\ \bibinfo {pages} {1655} (\bibinfo {year} {2017})},\ \Eprint
  {http://arxiv.org/abs/1610.09692} {arXiv:1610.09692 [hep-th]} \BibitemShut
  {NoStop}%
\bibitem [{\citenamefont {Carmona}\ \emph
  {et~al.}(2018{\natexlab{a}})\citenamefont {Carmona}, \citenamefont {Cortes},\
  and\ \citenamefont {Relancio}}]{Carmona:2017oit}%
  \BibitemOpen
  \bibfield  {author} {\bibinfo {author} {\bibfnamefont {J.~M.}\ \bibnamefont
  {Carmona}}, \bibinfo {author} {\bibfnamefont {J.~L.}\ \bibnamefont {Cortes}},
  \ and\ \bibinfo {author} {\bibfnamefont {J.~J.}\ \bibnamefont {Relancio}},\
  }\href {\doibase 10.1088/1361-6382/aa9ef8} {\bibfield  {journal} {\bibinfo
  {journal} {Class. Quant. Grav.}\ }\textbf {\bibinfo {volume} {35}},\ \bibinfo
  {pages} {025014} (\bibinfo {year} {2018}{\natexlab{a}})},\ \Eprint
  {http://arxiv.org/abs/1702.03669} {arXiv:1702.03669 [hep-th]} \BibitemShut
  {NoStop}%
\bibitem [{\citenamefont {Carmona}\ \emph
  {et~al.}(2018{\natexlab{b}})\citenamefont {Carmona}, \citenamefont
  {Cort\'es},\ and\ \citenamefont {Relancio}}]{Carmona:2018xwm}%
  \BibitemOpen
  \bibfield  {author} {\bibinfo {author} {\bibfnamefont {J.~M.}\ \bibnamefont
  {Carmona}}, \bibinfo {author} {\bibfnamefont {J.~L.}\ \bibnamefont
  {Cort\'es}}, \ and\ \bibinfo {author} {\bibfnamefont {J.~J.}\ \bibnamefont
  {Relancio}},\ }\href {\doibase 10.3390/sym10070231} {\bibfield  {journal}
  {\bibinfo  {journal} {Symmetry}\ }\textbf {\bibinfo {volume} {10}},\ \bibinfo
  {pages} {231} (\bibinfo {year} {2018}{\natexlab{b}})},\ \Eprint
  {http://arxiv.org/abs/1806.01725} {arXiv:1806.01725 [hep-th]} \BibitemShut
  {NoStop}%
\bibitem [{\citenamefont {Carmona}\ \emph
  {et~al.}(2019{\natexlab{b}})\citenamefont {Carmona}, \citenamefont {Cortes},\
  and\ \citenamefont {Relancio}}]{Carmona:2019oph}%
  \BibitemOpen
  \bibfield  {author} {\bibinfo {author} {\bibfnamefont {J.~M.}\ \bibnamefont
  {Carmona}}, \bibinfo {author} {\bibfnamefont {J.~L.}\ \bibnamefont {Cortes}},
  \ and\ \bibinfo {author} {\bibfnamefont {J.~J.}\ \bibnamefont {Relancio}},\
  }\href {\doibase 10.3390/sym11111401} {\bibfield  {journal} {\bibinfo
  {journal} {Symmetry}\ }\textbf {\bibinfo {volume} {11}},\ \bibinfo {pages}
  {1401} (\bibinfo {year} {2019}{\natexlab{b}})},\ \Eprint
  {http://arxiv.org/abs/1911.12700} {arXiv:1911.12700 [hep-th]} \BibitemShut
  {NoStop}%
\bibitem [{\citenamefont {Frattulillo}(2021)}]{frattulillo}%
  \BibitemOpen
  \bibfield  {author} {\bibinfo {author} {\bibfnamefont {D.}~\bibnamefont
  {Frattulillo}},\ }\href
  {https://www.youtube.com/watch?v=twAMuqn5C-U&list=PLUX8Mk7mqLPKmlEE4w2qfYg4km__j0I1U&index=3}
  {\enquote {\bibinfo {title} {Planck scale deformed realtivistic
  transformations in curved spacetime},}\ } (\bibinfo {year} {2021}),\ \bibinfo
  {note} {talk given at Corfu Summer Institute for the Second Annual Conference
  of COST CA18108, Greece}\BibitemShut {NoStop}%
\bibitem [{\citenamefont {Kowalski-Glikman}\ and\ \citenamefont
  {Nowak}(2002)}]{KowalskiGlikman:2002we}%
  \BibitemOpen
  \bibfield  {author} {\bibinfo {author} {\bibfnamefont {J.}~\bibnamefont
  {Kowalski-Glikman}}\ and\ \bibinfo {author} {\bibfnamefont {S.}~\bibnamefont
  {Nowak}},\ }\href {\doibase 10.1016/S0370-2693(02)02063-4} {\bibfield
  {journal} {\bibinfo  {journal} {Phys. Lett.}\ }\textbf {\bibinfo {volume}
  {B539}},\ \bibinfo {pages} {126} (\bibinfo {year} {2002})},\ \Eprint
  {http://arxiv.org/abs/hep-th/0203040} {arXiv:hep-th/0203040 [hep-th]}
  \BibitemShut {NoStop}%
\bibitem [{\citenamefont {Amelino-Camelia}\ \emph
  {et~al.}(2011{\natexlab{b}})\citenamefont {Amelino-Camelia}, \citenamefont
  {Freidel}, \citenamefont {Kowalski-Glikman},\ and\ \citenamefont
  {Smolin}}]{AmelinoCamelia:2011bm}%
  \BibitemOpen
  \bibfield  {author} {\bibinfo {author} {\bibfnamefont {G.}~\bibnamefont
  {Amelino-Camelia}}, \bibinfo {author} {\bibfnamefont {L.}~\bibnamefont
  {Freidel}}, \bibinfo {author} {\bibfnamefont {J.}~\bibnamefont
  {Kowalski-Glikman}}, \ and\ \bibinfo {author} {\bibfnamefont
  {L.}~\bibnamefont {Smolin}},\ }\href {\doibase 10.1103/PhysRevD.84.084010}
  {\bibfield  {journal} {\bibinfo  {journal} {Phys. Rev.}\ }\textbf {\bibinfo
  {volume} {D84}},\ \bibinfo {pages} {084010} (\bibinfo {year}
  {2011}{\natexlab{b}})},\ \Eprint {http://arxiv.org/abs/1101.0931}
  {arXiv:1101.0931 [hep-th]} \BibitemShut {NoStop}%
\bibitem [{\citenamefont {Magueijo}\ and\ \citenamefont
  {Smolin}(2002)}]{Magueijo:2001cr}%
  \BibitemOpen
  \bibfield  {author} {\bibinfo {author} {\bibfnamefont {J.}~\bibnamefont
  {Magueijo}}\ and\ \bibinfo {author} {\bibfnamefont {L.}~\bibnamefont
  {Smolin}},\ }\href {\doibase 10.1103/PhysRevLett.88.190403} {\bibfield
  {journal} {\bibinfo  {journal} {Phys. Rev. Lett.}\ }\textbf {\bibinfo
  {volume} {88}},\ \bibinfo {pages} {190403} (\bibinfo {year} {2002})},\
  \Eprint {http://arxiv.org/abs/hep-th/0112090} {arXiv:hep-th/0112090}
  \BibitemShut {NoStop}%
\bibitem [{\citenamefont {Carmona}\ \emph
  {et~al.}(2020{\natexlab{b}})\citenamefont {Carmona}, \citenamefont
  {Cortés},\ and\ \citenamefont {Relancio}}]{Carmona:2019vsh}%
  \BibitemOpen
  \bibfield  {author} {\bibinfo {author} {\bibfnamefont {J.~M.}\ \bibnamefont
  {Carmona}}, \bibinfo {author} {\bibfnamefont {J.~L.}\ \bibnamefont
  {Cortés}}, \ and\ \bibinfo {author} {\bibfnamefont {J.~J.}\ \bibnamefont
  {Relancio}},\ }\href {\doibase 10.1103/PhysRevD.101.044057} {\bibfield
  {journal} {\bibinfo  {journal} {Phys. Rev.}\ }\textbf {\bibinfo {volume}
  {D101}},\ \bibinfo {pages} {044057} (\bibinfo {year} {2020}{\natexlab{b}})},\
  \Eprint {http://arxiv.org/abs/1912.12885} {arXiv:1912.12885 [hep-th]}
  \BibitemShut {NoStop}%
\bibitem [{\citenamefont {Carmona}\ \emph {et~al.}(2012)\citenamefont
  {Carmona}, \citenamefont {Cortes},\ and\ \citenamefont
  {Mercati}}]{Carmona:2012un}%
  \BibitemOpen
  \bibfield  {author} {\bibinfo {author} {\bibfnamefont {J.~M.}\ \bibnamefont
  {Carmona}}, \bibinfo {author} {\bibfnamefont {J.~L.}\ \bibnamefont {Cortes}},
  \ and\ \bibinfo {author} {\bibfnamefont {F.}~\bibnamefont {Mercati}},\ }\href
  {\doibase 10.1103/PhysRevD.86.084032} {\bibfield  {journal} {\bibinfo
  {journal} {Phys. Rev. D}\ }\textbf {\bibinfo {volume} {86}},\ \bibinfo
  {pages} {084032} (\bibinfo {year} {2012})},\ \Eprint
  {http://arxiv.org/abs/1206.5961} {arXiv:1206.5961 [hep-th]} \BibitemShut
  {NoStop}%
\bibitem [{\citenamefont {Arzano}\ \emph {et~al.}(2019)\citenamefont {Arzano},
  \citenamefont {Kowalski-Glikman},\ and\ \citenamefont
  {Wislicki}}]{Arzano:2019toz}%
  \BibitemOpen
  \bibfield  {author} {\bibinfo {author} {\bibfnamefont {M.}~\bibnamefont
  {Arzano}}, \bibinfo {author} {\bibfnamefont {J.}~\bibnamefont
  {Kowalski-Glikman}}, \ and\ \bibinfo {author} {\bibfnamefont
  {W.}~\bibnamefont {Wislicki}},\ }\href {\doibase
  10.1016/j.physletb.2019.05.025} {\bibfield  {journal} {\bibinfo  {journal}
  {Phys. Lett. B}\ }\textbf {\bibinfo {volume} {794}},\ \bibinfo {pages} {41}
  (\bibinfo {year} {2019})},\ \Eprint {http://arxiv.org/abs/1904.06754}
  {arXiv:1904.06754 [hep-ph]} \BibitemShut {NoStop}%
\bibitem [{\citenamefont {Lobo}\ and\ \citenamefont
  {Pfeifer}(2021)}]{Lobo:2020qoa}%
  \BibitemOpen
  \bibfield  {author} {\bibinfo {author} {\bibfnamefont {I.~P.}\ \bibnamefont
  {Lobo}}\ and\ \bibinfo {author} {\bibfnamefont {C.}~\bibnamefont {Pfeifer}},\
  }\href {\doibase 10.1103/PhysRevD.103.106025} {\bibfield  {journal} {\bibinfo
   {journal} {Phys. Rev. D}\ }\textbf {\bibinfo {volume} {103}},\ \bibinfo
  {pages} {106025} (\bibinfo {year} {2021})},\ \Eprint
  {http://arxiv.org/abs/2011.10069} {arXiv:2011.10069 [hep-ph]} \BibitemShut
  {NoStop}%
\bibitem [{\citenamefont {Lobo}\ \emph {et~al.}(2021)\citenamefont {Lobo},
  \citenamefont {Pfeifer}, \citenamefont {Morais}, \citenamefont {Batista},\
  and\ \citenamefont {Bezerra}}]{Lobo:2021yem}%
  \BibitemOpen
  \bibfield  {author} {\bibinfo {author} {\bibfnamefont {I.~P.}\ \bibnamefont
  {Lobo}}, \bibinfo {author} {\bibfnamefont {C.}~\bibnamefont {Pfeifer}},
  \bibinfo {author} {\bibfnamefont {P.~H.}\ \bibnamefont {Morais}}, \bibinfo
  {author} {\bibfnamefont {R.~A.}\ \bibnamefont {Batista}}, \ and\ \bibinfo
  {author} {\bibfnamefont {V.~B.}\ \bibnamefont {Bezerra}},\ }\href@noop {} {\
  (\bibinfo {year} {2021})},\ \Eprint {http://arxiv.org/abs/2112.12172}
  {arXiv:2112.12172 [hep-ph]} \BibitemShut {NoStop}%
\bibitem [{\citenamefont {Carmona}\ \emph
  {et~al.}(2018{\natexlab{c}})\citenamefont {Carmona}, \citenamefont {Cortes},\
  and\ \citenamefont {Relancio}}]{Carmona:2017cry}%
  \BibitemOpen
  \bibfield  {author} {\bibinfo {author} {\bibfnamefont {J.~M.}\ \bibnamefont
  {Carmona}}, \bibinfo {author} {\bibfnamefont {J.~L.}\ \bibnamefont {Cortes}},
  \ and\ \bibinfo {author} {\bibfnamefont {J.~J.}\ \bibnamefont {Relancio}},\
  }\href {\doibase 10.1103/PhysRevD.97.064025} {\bibfield  {journal} {\bibinfo
  {journal} {Phys. Rev. D}\ }\textbf {\bibinfo {volume} {97}},\ \bibinfo
  {pages} {064025} (\bibinfo {year} {2018}{\natexlab{c}})},\ \Eprint
  {http://arxiv.org/abs/1711.08403} {arXiv:1711.08403 [hep-th]} \BibitemShut
  {NoStop}%
\bibitem [{\citenamefont {Gubitosi}\ and\ \citenamefont
  {Heefer}(2019)}]{Gubitosi:2019ymi}%
  \BibitemOpen
  \bibfield  {author} {\bibinfo {author} {\bibfnamefont {G.}~\bibnamefont
  {Gubitosi}}\ and\ \bibinfo {author} {\bibfnamefont {S.}~\bibnamefont
  {Heefer}},\ }\href {\doibase 10.1103/PhysRevD.99.086019} {\bibfield
  {journal} {\bibinfo  {journal} {Phys. Rev. D}\ }\textbf {\bibinfo {volume}
  {99}},\ \bibinfo {pages} {086019} (\bibinfo {year} {2019})},\ \Eprint
  {http://arxiv.org/abs/1903.04593} {arXiv:1903.04593 [gr-qc]} \BibitemShut
  {NoStop}%
\bibitem [{\citenamefont {Carmona}\ \emph {et~al.}(2021)\citenamefont
  {Carmona}, \citenamefont {Cort\'es},\ and\ \citenamefont
  {Relancio}}]{Carmona:2021gbg}%
  \BibitemOpen
  \bibfield  {author} {\bibinfo {author} {\bibfnamefont {J.~M.}\ \bibnamefont
  {Carmona}}, \bibinfo {author} {\bibfnamefont {J.~L.}\ \bibnamefont
  {Cort\'es}}, \ and\ \bibinfo {author} {\bibfnamefont {J.~J.}\ \bibnamefont
  {Relancio}},\ }\href {\doibase 10.3390/universe7040099} {\bibfield  {journal}
  {\bibinfo  {journal} {Universe}\ }\textbf {\bibinfo {volume} {7}},\ \bibinfo
  {pages} {99} (\bibinfo {year} {2021})},\ \Eprint
  {http://arxiv.org/abs/2104.07336} {arXiv:2104.07336 [gr-qc]} \BibitemShut
  {NoStop}%
\bibitem [{\citenamefont {Bevilacqua}\ and\ \citenamefont
  {Kowalski-Glikman}(2022)}]{Bevilacqua:2022fnc}%
  \BibitemOpen
  \bibfield  {author} {\bibinfo {author} {\bibfnamefont {A.}~\bibnamefont
  {Bevilacqua}}\ and\ \bibinfo {author} {\bibfnamefont {J.}~\bibnamefont
  {Kowalski-Glikman}},\ }in\ \href@noop {} {\emph {\bibinfo {booktitle} {{1st
  COST CA18108 First Training School~}}}}\ (\bibinfo {year} {2022})\ \Eprint
  {http://arxiv.org/abs/2203.04091} {arXiv:2203.04091 [physics.gen-ph]}
  \BibitemShut {NoStop}%
\end{thebibliography}%


%

\end{document}